\pdfoutput=1

\documentclass{bjour}
\usepackage{graphicx}
\usepackage{bjourps}

\usepackage{amsmath}
\usepackage{mathrsfs}
\DeclareMathAlphabet{\mathpzc}{OT1}{pzc}{m}{it}

\usepackage[superscript]{cite}

\def \piByTwo{\mbox{$\frac{\pi}{2}$}}
\def \piBy2{\mbox{$\frac{\pi}{2}$}}

\def \dltEff{\mbox{$\delta_{\text{eff}}$}}

\def \B0{\mbox{$\bf B_0$}}
\def \Bone{\mbox{$\bf B_1$}}

\def \Sqrt2{\mbox{$ {\sqrt{2}} $}}
\def \Sqrt2inv{\mbox{$ \frac{1}{\sqrt{2}} $}}

\def \c13{\mbox{$^{13}$C}}
\def \n15{\mbox{$^{15}$N}}
\def \h1{\mbox{$^1$H}}
\def \li7{\mbox{$^7$Li}}

\def \w1s{\mbox{\it b \ }}

\def \paper1{[26]}

\begin{document}

\title{
{\bf Visualizing electromagnetic vacuum by MRI}
}

\authors{ 
Chandrika S Chandrashekar$^1$,
Annadanesh Shellikeri$^2$,
S Chandrashekar$^{3*}$,
Erika A Taylor$^4$, Deanne M Taylor$^{5,6,7}$
}


\affil{
1. Lincoln High School (class of 2018), 3838 Trojan Trail, Tallahassee,
Florida, 32311, USA \\
2. Aeropropulsion, Mechatronics and Energy Center, Florida State University,
2003 Levy Ave., Tallahassee, FL 32310, USA \\
3. National High Magnetic Field Laboratory (NHMFL) and Florida State University,
1800 E. Paul Dirac Drive, Tallahassee, Florida, 32310, USA \\
4. Department of Chemistry, Wesleyan University,52 Lawn Ave., Hall-Atwater Labs,
Middletown, Connecticut 06459, USA \\
5. Department of Pediatrics, Perelman School of Medicine,
University of Pennsylvania, Philadelphia, Pennsylvania, 19104, USA \\
6. Department of Biomedical and Health Informatics,
The Children's Hospital of Philadelphia, Philadelphia, PA 19041, USA\\
7. Department of Genetics, Rutgers University, Piscataway, NJ 08854, USA\\
{\bf ($^*$ Corresponding author)} \\
start:  {\bf 20150601}
posted: {\bf 20160912}
rvsn1:  {\bf 20161124}
rvsn2:  {\bf 20161208}
}

\abstract{
Based upon Maxwell's equations, it has long been established that oscillating
electromagnetic (EM) fields incident upon a metal surface decay exponentially
inside the conductor \cite{jackson,si0griffiths,si0ulaby}, leading to a virtual 
EM vacuum at sufficient depths.  Magnetic resonance imaging (MRI) utilizes
radiofrequency (r.f.) EM fields to produce images.  Here we present the first 
visualization of a virtual EM vacuum inside a bulk metal strip by MRI, amongst
several novel findings.

We uncover unexpected MRI intensity patterns arising from two orthogonal pairs
of faces of a metal strip, and derive formulae for their intensity ratios,
revealing {\em differing effective elemental volumes} (voxels) {\em underneath
these faces}.

Further, we furnish chemical shift imaging (CSI) results that discriminate
different {\em faces} (surfaces) of a metal block according to their distinct
nuclear magnetic resonance (NMR) chemical shifts, which holds much promise for
monitoring surface chemical reactions noninvasively.

Bulk metals are ubiquitous, and MRI is a premier noninvasive diagnostic tool.
Combining the two, the emerging field of bulk metal MRI can be expected to grow
in importance. The fundamental nature of results presented here may impact bulk 
metal MRI and CSI across many fields.
} 

\begin{article}

\section{Introduction}
\label{sxn0intrdxn}
MRI is a household name as a diagnostic tool in the medical field
\cite{vp,tey},
with an impressive resume in many other fields, including the
study of materials \cite{callaghan,blumch,bjj,rugar},
corrosion of metals, monitoring batteries and supercapacitors
\cite{britton2010,ilott2014}.

However, historically, MRI of bulk metals has been very rare, limited to
specialized cells using r.f. gradients (with limited control) \cite{si0grld},
instead of the magnetic field gradients employed in conventional MRI.
In other studies involving bulk metals, the MRI targeted the surrounding
dielectric (electrolyte in electrochemical and fuel cells with metallic
electrodes, or tissues with embedded metallic implants)
\cite{si0pines,si0camacho,si0bennet,si0olsen,si0alm,si0viano,si0shafiei,si0graf,
si0koch,si0jfz,si0garwood2011,si0garwood}.
Notwithstanding, these studies addressed issues that can cause distortions and
limit sensitivity of the MRI images, such as bulk magnetic susceptibility (BMS) 
effects and eddy currents (produced on bulk metal surface due to
gradient switching).

The dearth of mainstream bulk metal MRI is rooted in unique challenges posed by 
the physics of propagation of r.f. EM fields in bulk metals
(MRI employs r.f. pulses to generate the MR signal leading to the images).
All along, it has been known that the incident r.f. field decays rapidly and
exponentially inside the metallic conductor (Fig.\ref{figSI0dltaEff}), a
phenomenon known as skin effect \cite{jackson,si0griffiths,si0ulaby}.
The characteristic length of decay ($\delta$, the {\em skin depth}), typically
of the order of several microns (Eq.(\ref{eq:eq0sknDpth})),
characterizes the limited r.f. penetration into the metal. This in turn, results
in attenuated MR signal intensity for bulk metals \cite{si0pines,rangeet}.
Turning the tables, Bhattacharyya et. al.,\cite{rangeet} exploited the skin
effect to separate and quantify bulk and non-bulk metal NMR signals in Li ion
batteries to monitor the growth of dendritic metallic structures.
Subsequently, for bulk metal MRI, yet another impediment was correctly diagnosed
\cite{chandrashekar}.
It was found that the orientation of the bulk metal surface, relative to \Bone
(the r.f. magnetic field),
critically affected the outcome.
Using optimal alignment of the bulk metal (electrodes), relative to \Bone,
recent studies have successfully demonstrated and highlighted bulk metal MRI
albeit, primarily applied to batteries and electrochemical cells
\cite{chandrashekar,ilott,romanenko,britton2014,hjc2015}.

Though unanticipated at the time, the recent bulk metal MRI findings eased the
implementation of MRI of liquid electrolyte, by helping mitigate adverse effects
due to the metal in the vicinity of lithium, zinc and titanium electrodes,
yielding fresh insights
\cite{romanenko,britton2014,si0britton2013,si0klett,si0furo}.
Similar benefits may be expected to accrue for MRI based radiology of soft
tissues with embedded metallic implants (pacemeakers, prosthetics, dental
implants, etc.).
\\

Here, we present several key findings on bulk metal MRI and CSI:
\begin{itemize}
\item
During a systematic noninvasive thickness measurement of bulk metal strips
by MRI \cite{romanenko}
(section S\ref{sxnSI0thcknss}),
we come across unexpected regions of intensity, and assign them to two mutually 
orthogonal pairs of faces of the strip.

\item
To explain the peculiar ratios of intensities from these different regions in
bulk metal MRI and CSI, we derive formulae from first principles, unveling
a surprising underlying reason: 
{\em differing effective elemental volumes for these different regions}.

\item
In the process, the images enable a visualization of a virtual EM vacuum
inside the bulk metal via an {\em MRI tunnel}.

\item
Additionally, we demonstrate that the bulk metal CSI distinguishes different
faces (surfaces) of a metal block according to their distinct NMR chemical
shifts.

\end{itemize}

We attained these by employing three {\em phantoms} (samples)
{\bf P0, P1, P3}, depicted schematically in Fig.\ref{fig0phntms} and described
in Methods section \ref{subsxn0mthdsPhntms}.
All phantoms were derived from the same stock of 0.75 $mm$ thick lithium (Li)
strip. Phantom P3 is a super strip composed of three Li strips pressed together,
forming an {\em effective} single strip three times thicker than the
individual strips in phantoms P0 and P1.

The setup of phantoms, r.f. coil and the gradient assembly ensures that the
imaging directions $x,y,z$ fulfill the condition that 
$x \parallel a \parallel \Bone$ and  $z \parallel \B0$
(the static main magnetic field), with the possibility to reorient the phantom
about the $x$-axis; $a,b,c$  are the three sides of the strips.

Since all MRI and CSI images to follow were acquired with the given phantom's
$bc$ faces {\em normal} ($\perp$, {\em perpendicular}) to \Bone, these images
bear the imprint of having no contribution from these faces to the magnetic
resonance (MR) signal \cite{chandrashekar,ilott,romanenko,britton2014}, since
\Bone\ penetration into the metal is maximal when it is {\em parallel}
($\parallel$) to the metal surface, and minimal when $\perp$ metal surface
\cite{jackson,chandrashekar,ilott}.

For details on the MRI experiments, including the nomenclature, the reader is
referred to 
Methods section \ref{subsxn0mthdsMri}.

\section{MRI}
\label{sxn0mri}
Fig.\ref{fig0xy0yz} furnishes stackplots (intensity along the vertical axis) of
\li7 2d MRI (without slice selection) of phantom P3.
Panel (a) displays MRI($xy$). Panel (b) displays MRI($yz$).

It is straightforward to infer that the {\em walls} of high intensity regions in
either image emanate from $ac$ faces of the P3 strip
\cite{chandrashekar,ilott,romanenko}, as we did while measuring the
thickness of metal strips (Figs.\ref{fig0twoStrps} and \ref{fig0xy},
section S\ref{sxnSI0thcknss}).
In either image, contributions along the non imaged axis sum up to yield the
high intensity walls.

However, the unexpected intensity between the two $ac$ faces of the super strip,
in both the images, is perplexing.
The 2d MRI($xy$) in Fig.\ref{fig0xy0yz}a exhibits a low intensity {\em plateau}
spanning the walls.
The 2d MRI($yz$) in Fig.\ref{fig0xy0yz}b displays low intensity {\em ridges}  
bridging the walls.
\\

\subsection{Visualizing a virtual eletromagnetic vacuum by MRI}
To understand better these unexpected regions of intensity, we acquired
\li7 3d MRI($xyz$) of phantom P3, shown in Fig.\ref{fig0xyz}.
In addition to the $ac$ faces (separated along $y$),
the $ab$ faces (separated along $z$) are revealed for the first time.

As noted earlier, $bc$ faces (being $\perp$ to \Bone) are absent.
The hollow region in MRI($xyz$) arises from the skin depth phenomenon
\cite{rangeet,chandrashekar,ilott,jackson,si0griffiths,si0ulaby},
restricting the EM fields to effectively access only a limited
subsurface underneath the $ac$ and $ab$ faces
(section S\ref{sxnSI0subSurface} and Fig.\ref{figSI0dltaEff}).
The presence of faces $\parallel$ \Bone, coupled with the conspicuous absence
of faces $\perp$ \Bone, in combination with the hollow region, imparts the 3d
image an appearance of an {\em MRI tunnel}, supplying a compelling visualization
of a virtual {\em EM vacuum} in the interior of a metallic conductor
(hitherto depicted only schematically in literature
(for e.g. Ref.\cite{rangeet})).
\\

With the aid of 3d MRI in Fig.\ref{fig0xyz}, the intensity regions in
2d MRI($xy$) and 2d MRI($yz$) images of Fig.\ref{fig0xy0yz}, can be easily
interpreted as simply regions resulting respectively from projections
along $z$ and $x$ axes of the 3d image.
It is convincingly clear that the intensity between $ac$ faces
(either the plateau or the ridges), is due to the pair of $ab$ faces
of the superstrip P3.

Yet, the basis for the relative intensity values remains elusive at
this stage.

For the 2d MRI($xy$) in Fig.\ref{fig0xy0yz}a, it can be argued that,
for the $ac$ face the entire length of side $c=$7 $mm$
(Fig.\ref{fig0phntms}) contributes to the signal,
while for the $ab$ face, only a subsurface depth
$ \dltEff \approx$ 9.49 $\mu m$ contributes
(Eq.(\ref{eq:eq0sknDpth}), Eq.(\ref{eq:eq0dltaEff}),
section S\ref{sxnSI0subSurface} and Fig.\ref{figSI0dltaEff}).
This would lead to a ratio of the corresponding intensities, $S_{ac}/S_{ab}$,
to be of the order of $c/(2 \dltEff) \approx 368 $ (Fig.\ref{fig0xy0yz0sim}a),
in obvious and jarring disagreement with the observed ratio
(of maxima of $S_{ac}$ and $S_{ab}$) of 6.6.

For the 2d MRI($yz$) in Fig.\ref{fig0xy0yz}b, the expected intensity pattern in 
a stack plot would be one with equal intensities from $ab$ and $ac$ faces,
since they share the same side, $a$, along $x$ (non imaged) axis
(Fig.\ref{fig0xy0yz0sim}b).
This again, is in stark contrast with the observed ratio
(of maxima of $S_{ac}$ and $S_{ab}$) of 10.

For the MRI($xyz$), naively, uniform intensity would be expected from both $ab$ 
and $ac$ faces, resulting in a ratio of unity.  Instead, the observed ratio
(of maxima of $S_{ac}$ and $S_{ab}$) is found to be 3.8.
 
Thus, the MRI images bear peculiar intensity ratios from comfortably identified 
(from 3d MRI) regions of the bulk metal.
We will return to this topic later.
\\

\section{CSI}
\label{sxn0csi}
The \li7 NMR spectrum of phantom P3 (Fig.\ref{fig0csi} inset)
contains two distinct peaks in the Knight shift region for metallic \li7
(see Methods section \ref{subsxn0mthdsMri}),
centered at $\delta_1$= 256.4 and $\delta_2$= 266.3 ppm.
At first sight it might seem odd that a metallic strip, of regular geometry and
uniform density, that is entirely composed of identical Li atoms, gives
rise to two NMR peaks instead of the expected single peak.

To gain additional insight as to the spatial distribution of the Li metal
species with different NMR shifts, we turn to CSI, which combines an NMR
chemical shift (CS) dimension with one or more imaging(I) dimensions
\cite{callaghan,haacke,si0spnglr,si0kwf}.

The 2d CSI($y$) shown in Fig.\ref{figSI0csi} comprises of two bands separated
along $y$ located at $\delta_2$, along the CS dimension,
while a low intensity band spans them along $y$ at a CS of $\delta_1$, strongly
hinting at, the two bands (at $\delta_2$) being associated with $ac$ faces.

This observation called for adding an additional
imaging dimension along $z$, leading to 3d CSI($yz$), which is realized in
Fig.\ref{fig0csi},
where $y$ and $z$ are the imaging dimensions, accompanied by the CS dimension.
The bands separated along $z$, occur at $\delta_1$.
The bands separated along $y$, occur at $\delta_2$.
In conjunction with the 3d MRI image in Fig.\ref{fig0xyz},
it is evident that the pairs of bands at $\delta_1$ and $\delta_2$ arise from
$ab$ and $ac$ faces respectively, completing the spatio-chemical assignment.

These assignments readily carry over to 2d CSI($y$) in Fig.\ref{figSI0csi},
with the pair of bands at $\delta_2$ and the low ridge spanning them at 
$\delta_1$, being respectively identified with $ac$ and $ab$ faces.  Similarly,
in the NMR spectrum, the short and tall peaks
respectively at $\delta_{1,2}$ are assigned to $ab$ and $ac$ faces, consistent
with the reported \cite{rangeet,ilott} experiments and simulations.

That different types of faces of the bulk metal strip suffer different NMR
(Knight) shifts according to their orientations {\em relative} to \B0,
is consistent with previous observations and simulations \cite{rangeet,ilott},
and has been traced to bulk magnetic susceptibility (BMS) effect
\cite{rangeet,chandrashekar,ilott,hjc2015,hoffman,lina,hjc}.

Interestingly, the 3d CSI sheds new light on previous bulk metal NMR studies.
For instance, in an earlier study \cite{rangeet}, a similar shift
difference between NMR peaks was observed at $\parallel$ and $\perp$
orientations (relative to \B0) of the major faces of a thinner metal strip,
by carrying out two {\em separate} experiments.
Here, phantom P3 furnishes these two orientations in a {\em single} experiment, 
via $ac$ and $ab$ faces (Fig.\ref{fig0phntms}).
The present work provides physical insight into another previous \cite{ilott}
observation. It was found that the intensity of NMR peak arising from $ab$
faces, unlike that from the $ac$ faces, was invariant under rotation about
$z \parallel c \parallel B_0$ axis.
Our 3d MRI (Fig.\ref{fig0xyz}) and 3d CSI (Fig.\ref{fig0csi}), directly
demonstrate that such a rotation leaves the orientation of \Bone\ relative to
$ab$ face (but not the $ac$ face) the same (signal intensity from a given face
depends on its orientation relative to \Bone \cite{chandrashekar,ilott}).
Note that the shifts themselves remain unshifted since they depend on the
orientation of the faces relative to \B0, which does not change under this
rotation ($ac$ and $bc$ faces remain $\parallel \B0$, whilst $ab$ faces
remain $\perp \B0$).

Thus, bulk metal CSI supplies direct
evidence, that the bulk metal chemical (Knight) shifts resulting from BMS are
correlated with the differing orientations (relative to \B0) of different parts 
of the bulk metal.\\

Like for MRI, the basis for the ratio of intensities from the $ac$ and $ab$
faces ($S_{ac}/S_{ab} \approx$ 2.8), is not immediately intuitively obvious
and will be explored next.

\section{Intensity ratio formulae for bulk metal MRI and CSI}
\label{sxn0formulae}
%
%
The peculiar intensity ratios in MRI and CSI, of signals $S_{ab}$ and $S_{ac}$,
arising respectively from $ab$ and $ac$ faces of phantom P3
(Fig.\ref{fig0xy0yz}, sections \ref{sxn0mri} and \ref{sxn0csi}),
could be due to gradient switching involved in the MRI experiments (the
resultant eddy currents could be different for $ab$ and $ac$ faces).
However, as shown in section S\ref{sxnSI0mri2dNtnst}, this can be ruled out on
the basis of 2d MRI($yz$) and MRI($zy$) at mutually orthogonal orientations
(related by a rotation about $x \parallel a \parallel \Bone$),
shown in Fig.\ref{fig0p3hrzntl}.

And yet, it is possible to derive, from elementary considerations
and first principles, and arrive at expressions for the {\em ratios} of MRI
and CSI intensities from $ab$ and $ac$ faces.\\

For the 2d MRI($xy$), in Fig.\ref{fig0xy0yz}a, the signal intensity from the
$ab$ faces can be written as (see Eq.(\ref{eq:eq1dltaEff}))
\begin{equation}
 S_{ab} (x,y)\propto dx\ dy\ \int dz =  dx\ dy\ 2\delta_{\text{eff}}
\label{eq:eq0sabxy}
\end{equation}
with  $dx\ dy\ dz$ denoting {\em elemental} volume of the metal,
and \dltEff\ is the {\em effective} subsurface {\em depth} that would account
for the MR signal in the {\em absence} of \Bone\ decay
(see Eq.(\ref{eq:eq0dltaEff}) 
and Fig.\ref{figSI0dltaEff}).
Above, the integral over $z$, is replaced by \dltEff\, underneath the two $ab$
faces separated along $z$.

Similarly, for the signal intensity from {\em either} of the $ac$ faces,
\begin{equation}
 S_{ac} (x,y)\propto dx\ dy\ \int dz = dx\ \delta_{\text{eff}}\ c
\label{eq:eq0sacxy}
\end{equation}
since the subsurface now is $\perp y$.

Eq.(\ref{eq:eq0sabxy}) and Eq.(\ref{eq:eq0sacxy}), reveal {\em differing
effective elemental volumes}(voxels) underneath these faces:
\begin{equation}
dV_{\text{eff}}^{\text{ab}} = dx\ dy\ \delta_{\text{eff}}
\label{eq:eq0dVeffab}
\end{equation}
\begin{equation}
dV_{\text{eff}}^{\text{ac}} = dx\ \delta_{\text{eff}}\ dz 
\label{eq:eq0dVeffac}
\end{equation}
From Eq.(\ref{eq:eq0sabxy}) and Eq.(\ref{eq:eq0sacxy}),
\begin{equation}
 \frac{S_{ac}}{S_{ab}} =\frac{c}{2\Delta y}
\label{eq:eq0ratioxy}
\end{equation}
where we have replaced $dy$ by $\Delta y$, the resolution along
$y\ \parallel b$.
Consulting the Methods section \ref{subsxn0mthdsMri} and Fig.\ref{fig0phntms},
$c=7$ $mm$, $\Delta y$=0.25 $mm$ and Eq.(\ref{eq:eq0ratioxy}) yields a
calculated ratio of $S_{ac}/S_{ab}$= 14
(as illustrated in Fig.\ref{fig0xy0yz0drvd}a),
within an order of magnitude of the observed ratio (section \ref{sxn0mri},
Fig.\ref{fig0xy0yz}a) and a 25 fold improvement relative to the expected ratio
(Fig.\ref{fig0xy0yz0sim}a).

Also, Eq.(\ref{eq:eq0ratioxy}) reveals that $S_{ac}/S_{ab}$ increases with
increasing resolution along $y$, as shown in the three images in
Fig.\ref{figSI0td20td40td80},
with relative resolutions increasing by factors of 1, 2 and 4,
yielding calculated  $S_{ac} / S_{ab}$ ratios of 7, 14 and 28 respectively.
The corresponding observed ratios (of maxima of $S_{ac}$ and $S_{ab}$) of
3.3, 6.6, and 11.6, are within an order of magnitude of the calculated values.
Remarkably, these observed ratios themselves increase by factors of 1, 2 and
3.5, mimicking closely the corresponding factors of resolution increase.
\\

Continuing in the same vein, for the 2d MRI($yz$), in Fig.\ref{fig0xy0yz}b,
\begin{equation}
 S_{ab} (y,z) \propto dy\ dz \int dx = a\ dy\ \delta_{\text{eff}}
\label{eq:eq0sabyz}
\end{equation}
while,
\begin{equation}
 S_{ac} (y,z) \propto dy\ dz \int dx = a\ \delta_{\text{eff}}\ dz
\label{eq:eq0sacyz}
\end{equation}
leading to
\begin{equation}
 \frac{S_{ac}}{S_{ab}}= \frac{\Delta z}{\Delta y}
\label{eq:eq0ratioyz}
\end{equation}
once again, replacing $dy,\ dz$ by $\Delta y,\ \Delta z$, the respective
resolutions along $y,\ z$.
Using the values of $\Delta y, \Delta z$= 0.0357, 1 $mm$
in Eq.(\ref{eq:eq0ratioyz}), ensues
a calculated ratio of $S_{ac} / S_{ab} \approx$ 28
(as illustrated in Fig.\ref{fig0xy0yz0drvd}b),
within an order of magnitude of the observed ratio (section \ref{sxn0mri},
Fig.\ref{fig0xy0yz}b), and 3.5 fold better than the expected ratio
(Fig.\ref{fig0xy0yz0sim}b). More importantly, the expected intensity pattern is 
even {\em qualitatively} (visually) different from the experiment, unlike the
derived pattern.

Similarly, for the 2d MRI($zy$) in Fig.\ref{fig0p3hrzntl}b, of phantom P3 in
{\em horizontal} orientation, it can be easily shown that, 
\begin{equation}
 \frac{S_{ac}}{S_{ab}}= \frac{\Delta y}{\Delta z}
\label{eq:eq0ratiozy}
\end{equation}
Using the values of $\Delta z, \Delta y$= 0.0357, 1 $mm$
in Eq.(\ref{eq:eq0ratiozy}), results in a calculated ratio of
$S_{ac} / S_{ab} \approx$ 28, within an order of magnitude of the observed
ratio (section S\ref{sxnSI0mri2dNtnst}).\\

Proceeding along the same lines, for the MRI($xyz$) in Fig.\ref{fig0xyz},
\begin{equation}
 S_{ab} (x,y,z) \propto dx\ dy\ dz =  dx\ dy\ \delta_{\text{eff}}
\label{eq:eq0sabxyz}
\end{equation}
while,
\begin{equation}
 S_{ac} (x,y,z) \propto dx\ dy\ dz =  dx\ \delta_{\text{eff}}\ dz
\label{eq:eq0sacxyz}
\end{equation}
As usual by now, replacing $dy,\ dz$ by $\Delta y,\ \Delta z$, the respective
resolutions along $y,\ z$, we obtain again Eq.(\ref{eq:eq0ratioyz}).
Consulting the Methods section \ref{subsxn0mthdsMri}, 
$\Delta y, \Delta z$= 0.25, 1 $mm$, respectively.  Using these values in 
Eq.(\ref{eq:eq0ratioyz}) yields a calculated ratio of $S_{ac} / S_{ab}$= 4,
within an order of magnitude of the observed ratio
(section \ref{sxn0mri}).

Fig.\ref{fig0xyzSlc25} shows (in stack plot representation, with vertical axis
denoting intensity) $xy$ slices (along $z$) from the 3d MRI($xyz$).
The central slice contains no signal between the walls of intensity (from $ac$
faces) as expected.
However, the slice from the top $ab$ face exhibits a {\em plateau} of intensity
between the $ac$ faces, visually demonstrating that $S_{ab} \neq S_{ac}$ in the 
non-hollow regions of the 3d image, in place of the expected uniform intensity.
\\

Similarly, for the 3d CSI($yz$) in Fig.\ref{fig0csi}, it can be shown that
the ratio $S_{ac} / S_{ab}$ is given by Eq.(\ref{eq:eq0ratioyz}), which along
with the relevant experimental parameters for this image,
yields a calculated value of 2, within an order of magnitude of the observed
ratio (of maxima of $S_{ac}$ and $S_{ab}$) of 2.8.

On the other hand, for the 2d CSI($y$), in Fig.\ref{figSI0csi}, it can be shown 
that the ratio $S_{ac} / S_{ab}$ is given by Eq.(\ref{eq:eq0ratioxy}), from
which we obtain a calculated value of 14, using the experimental parameters in
section \ref{subsxn0mthdsMri}.  The measured ratio (of maxima of $S_{ac}$ and
$S_{ab}$) of 9.5, is again within an order of magnitude of the calculated value.
\\

Thus, the $S_{ac} / S_{ab}$ ratios calculated from
Eqs.(\ref{eq:eq0ratioxy}), (\ref{eq:eq0ratioyz}) and (\ref{eq:eq0ratiozy}),
agree with the observed values within an order of magnitude
for 2d MRI($xy$), 2d MRI($yz$), 2d MRI($zy$), 3d MRI($xyz$), 3d CSI($yz$) and
2d CSI($y$).
In fact, discrepancies between observed and derived $S_{ac}/S_{ab}$ ratios range
only by factors of 0.7 to 2.8 across various MRI and CSI images
(see Table.\ref{table:tbl0ratio}).
More importantly, the derived patterns {\em resemble} the observed patterns,
unlike the expected patterns, which differ even visually from the observed
patterns (for e.g., see
Figs.\ref{fig0xy0yz}, \ref{fig0xy0yz0sim} and \ref{fig0xy0yz0drvd}).

\begin{table}[h]
 \caption{
  The observed, derived (from
  Eqs.(\ref{eq:eq0ratioxy}), (\ref{eq:eq0ratioyz}), (\ref{eq:eq0ratiozy})),
  and expected (from skin depth arguments alone) $S_{ac} / S_{ab}$ ratios.
 }
 \begin{tabular}{|c| c c c|}
  \hline
    & & $S_{ac} / S_{ab}$    &  \\
  \hline
   Experiment & Observed & Derived & Expected \\
\hline
MRI($xy$) & & & \\
Fig.\ref{figSI0td20td40td80}a &  3.3 &  7 & 368 \\
Fig.\ref{figSI0td20td40td80}b &  6.6 & 14 & 368 \\
Fig.\ref{figSI0td20td40td80}c & 11.6 & 28 & 368 \\
 & & & \\
MRI($yz$)    & 10  & 28  &   1 \\
MRI($zy$)    & 10  & 28  &   1 \\
MRI($xyz$)   & 3.8 &  4  &   1 \\
2d CSI($y$)  & 9.5 & 14  & 368 \\
3d CSI($yz$) & 2.8 &  2  &   1 \\
\hline
 \end{tabular}
 \label{table:tbl0ratio}
\end{table}

In summary, the formulae unveil the underlying reason for the significant
departure of observed $S_{ac}/S_{ab}$ from expected values:
{\em differing effective elemental volumes underneath these faces},
as revealed by Eqs.(\ref{eq:eq0dVeffab}) and (\ref{eq:eq0dVeffac}).
The derived patterns bear closer resemblance to
experiment, than what is expected from skin depth consideratons alone, or from
conventional specifications of the voxel= $\Delta x \Delta y \Delta z$ (see for
e.g.  Figs.\ref{fig0xy0yz}, \ref{fig0xy0yz0sim} and \ref{fig0xy0yz0drvd}).

On a practical note, these formulae can guide experimental strategies to
relatively enhance MRI and CSI signals from different regions of the bulk metal.

\section{Conclusions}
\label{sxn0cnclusn}
In conclusion, the unexpected findings presented here may impact bulk metal MRI 
and CSI studies in general,
via fresh insights
for data collection, analysis and interpretation.
The bulk metal MRI and CSI
(correlating different bulk metal surfaces with distinct chemical shifts)
results in this study have the noninvasive diagnostic potential in other fields
such as
structure  of metals and alloys \cite{rdrgz,flyn},
metallurgy (metal fatigue, fracture, strain)
\cite{mtllurg,bppag,mvrkks},
catalysis 
\cite{zhong,yuan},
bulk metal surface science and surface chemistry
\cite{tlptr,whttn,rgg},
metallic medical implants, dielectric MRI in the vicinity of bulk metals
etc.
(section \ref{sxn0intrdxn}).

The findings may also lead to as yet unforeseen applications
(section \ref{sxn0intrdxn}) since,
(i) they are of a fundamental nature,
(ii) there are no inherent limitations to the approach employed
(scalability, different metals, systems other than batteries, etc., are all
possible),
(iii) the study utilizes only standard MRI tools (hardware, pulse sequences,
data acquisition and processing), ensuring ease of implementation and
reproducibility. Thus it is likely to benefit from advances made in the
mainstream (medical) MRI field.

\section{Methods}
\label{sxn0mthds}
\subsection{Phantoms}
\label{subsxn0mthdsPhntms}
All phantoms were assembled and sealed in an argon filled glove box.
All three phantoms, P0, P1 and P3, shown in Fig.\ref{fig0phntms} 
(of dimensions $a \times b \times c$),
were derived from a (0.75 $mm$ thick) stock Li strip (Alfa Aesar 99.9\%).
The Li strips were mounted on a 2.3 $mm$ thick teflon strip and the resulting
sandwich bound together with Kapton tape.
Each phantom was placed in a flat bottom glass tube (9.75 $mm$ inner diameter
(I.D.), 11.5 $mm$ outer diameter (O.D.) and  5 $cm$ long), with
the longest side, $a$, $\parallel$ to the tube axis and to the axis of the
home built horizontal loop gap resonator (LGR)  r.f. coil (32 $mm$ long, 15 $mm$
O.D.), thus guaranteeing $\Bone \parallel a$.
The phantom containing glass tubes were wrapped with Scotch tape
to snugly fit into the r.f. coil.

In  Fig.\ref{fig0phntms}, $x,y,z$ specify the imaging (gradient) directions,
with $z \parallel \B0$ (the main magnetic field).
For our horizontal LGR r.f. coil (the MR resonator) and the gradient assembly
system, $\Bone \parallel x$, resulting in $\Bone \parallel x \parallel a$.
\begin{itemize}
\item {\bf P0}: Pair of Li strips separated by a teflon strip;
for each Li strip,
$a \times b \times c=$ 20 x 0.75 x 7 $mm^3$.
\item
{\bf P1}: Single Li strip.
$a \times b \times c=$ 15 x 0.75 x 7 $mm^3$.
\item
{\bf P3}: Three  Li strips pressed together to yield a single
composite super strip.
$a \times b \times c=$ 15 x 2.25 x 7 $mm^3$.
\end{itemize}

\subsection{MRI and CSI}
\label{subsxn0mthdsMri}
Magnetic resonance experiments were conducted on a $B_0$=21$T$ magnet
(corresponding to \li7 Larmor frequency of 350 MHz)
operating under Bruker Avance III system with Topspin spectrometer control and
data acquisition, and equipped with a triple axes ($x,y,z$) gradient amplifier
assembly, using a multinulcear MRI probe
(for a triple axes 63 $mm$ I.D. gradient stack by Resonance Research Inc.), 
employing the LGR r.f. coil (resonating at 350 MHz) desribed above.

The MRI and CSI data were acquired using spin-echo imaging pulse sequence
without slice selection \cite{callaghan,haacke} (yielding sum total of signal
conributions from the non imaged dimensions).
Frequency encoding gradient was
employed for the directly detected dimension and phase encoding gradients for
the indirect dimensions \cite{callaghan,haacke}.
The CSI experiments were carried out with the NMR chemical shift as the directly
detected dimension, with phase encoding gradients along the indirectly detected 
imaging dimensions.
The r.f. pulses were applied at a carrier frequency of 261 ppm (to excite the
metallic \li7 nuclear spins in the Knight shift region
\cite{rangeet,abragam}),
typically with a strength of 12.5 $kHz$, with a recycle (relaxation) delay of
0.5 $s$.
The gradient dephasing delay and phase encoding gradient duration were 0.5 $ms$.

Throughout this manuscript,
the first axis label ($x,y,z$) describing an MRI experiment stands for frequency
encoding dimension and the remaining ones correspond to phase econded
dimensions.   For e.g., MRI($xyz$) implies frequency encoding along $x$ axis,
and phase encoding along the remaining directions.

$G_x,G_y,G_z$ and $N_x,N_y,N_z$ denote respectively the gradient strengths in
units of $T/m$ and number of data points in $k$-space ($^*$ denoting complex
number of points acquired in quadrature) \cite{callaghan,haacke}, along
$x,y,z$ axes.
$L_x,L_y,L_z$ and $\Delta x,\Delta y,\Delta z$ are respectively the resultant
nominal field of view (FOV) and resolution, in units of $mm$, along $x,y,z$ axes
\cite{callaghan,haacke}.

Also, $n$ is the number of transients accumulated for signal averaging and
$SW$ is the spectral width (in units of $kHz$) for the directly detected
dimension in MRI and CSI.
\\ \\
{\bf 1d MRI({\em y}):} \\
$n=64,\ SW=50$ \\
$G_y=0.42,\ N_y^*=200,\ L_y=7.143,\ \Delta y= 0.0357$
\\
{\bf 2d MRI({\em xy}):} \\
$n=32,\ SW=100$ \\
$G_x=0.24,\ N_x^*=200,\ L_x=25,\ \Delta x= 0.125$ \\
$L_y=10$ \\
(1) $G_y=0.12,\ N_y=20,\  \Delta y= 0.500$
     (Figs.\ref{fig0twoStrps}a,\ref{figSI0td20td40td80}a) \\
(2) $G_y=0.24,\ N_y=40,\  \Delta y= 0.250$
     (Figs.\ref{fig0twoStrps}b,\ref{fig0xy}b,\ref{fig0xy0yz}a,
     \ref{figSI0td20td40td80}b) \\
(3) $G_y=0.48,\ N_y=80,\  \Delta y= 0.125$ (Fig.\ref{figSI0td20td40td80}c) \\
{\bf 2d MRI({\em yz}):} \\
$n=32$  
$SW=50$ \\
$G_y=0.42,\ N_y^*=200,\ L_y=7.143,\ \Delta y= 0.0357$ \\
$G_z=0.06,\  N_z=16,\    L_z=16,\ \Delta z= 1.000$
\\
{\bf 2d MRI({\em zy}):} \\
$n=32,\ SW=50$ \\
$G_z=0.42,\ N_z^*=200,\ L_z=7.143,\ \Delta z= 0.0357$ \\
$G_y=0.06,\  N_y=16,\    L_y=16,\ \Delta y= 1.000$
\\
{\bf MRI({\em xyz}):} \\
$n=16,\ SW=100$ \\ 
$G_x=0.24,\ N_x^*=200,\ L_x=25,\ \Delta x= 0.125$ \\
$G_y=0.24,\ N_y=40,\    L_y=10,\ \Delta y= 0.250$ \\
$G_z=0.06,\  N_z=16,\    L_z=16,\ \Delta z= 1.000$
\\
{\bf 2d CSI({\em y}):} \\
$n=8,\  SW=100$, number of data points(complex)=$1024$ \\ 
$G_y=0.24,\ N_y=40,\    L_y=10,\ \Delta y= 0.250$
\\
{\bf 3d CSI({\em yz}):} \\
$n=24,\ SW=100$, number of data points(complex)=$1024$ \\ 
$G_y=0.12,\ N_y=20,\    L_y=10,\ \Delta y= 0.500$ \\
$G_z=0.06,\  N_z=16,\    L_z=16,\ \Delta z= 1.000$ \\
\\
All data were processed in Bruker's Topspin, with one zero fill prior to complex
fast Fourier Transform (FFT) along each dimension either without any window
function or with sine-bell window function.
All data were 'normalized' (to $\approx 10$, for plotting convenience) to aid
comparing {\em relative} intensities from different regions {\em within} a given
image.
For the purpose of determining the ratios of signal intensities associated
with different regions of the bulk metal, the intensity values were measured
directly from the processed images either in Topspin or Matlab (for e.g.,
'datatip' utility in Matlab, yields the coordinates and the 'value' (intensity) 
of a data point by clicking on it, in 1d, 2d and 3d plots).


{\bf AUTHOR CONTRIBUTIONS}\\
CSC recognized and analyzed anomaly between observed and expected intensity
    patterns in the MRI and CSI images,
    prepared figures, analyzed data, interpreted and discussed results, helped
    write and proof read manuscript, carried out literature search.\\
AS  prepared the phantoms, discussed results, helped write the manuscript. \\
EAT discussed results, helped write the manuscript.\\
DMT discussed results, helped write the manuscript.\\
SC  conceptualized the project, designed phantoms, designed and conducted
    experiments, developed analytical tools, organized and analyzed data,
    interpreted results, derived intensity ratio equations, helped with figures,
    and wrote the manuscript.\\
All authors reviewed the manuscript. \\ \\
{\bf Additional Information} \\
{\bf Competing financial interests} \\
The authors declare no competing financial interests.


\end{article}


\begin{figure}
\caption{
{\bf Skin depth $\delta$, and {\em effective} subsurface depth
$\delta_{\text{eff}}$.}\\
{\bf (a)}
Exponential decay of magnitude of \Bone($y$), as a function of the depth from
the metal surface (see Eq.(\ref{eq:eq0b1exp}) in
section S\ref{sxnSI0subSurface}).\\
{\bf (b)}
{\em Effective} subsurface depth $\delta_{\text{eff}}$, {\em without}
\Bone decay, that would account for the MR signal (see Eq.(\ref{eq:eq0dltaEff}) 
in section S\ref{sxnSI0subSurface}).
}
\includegraphics[width=8cm]{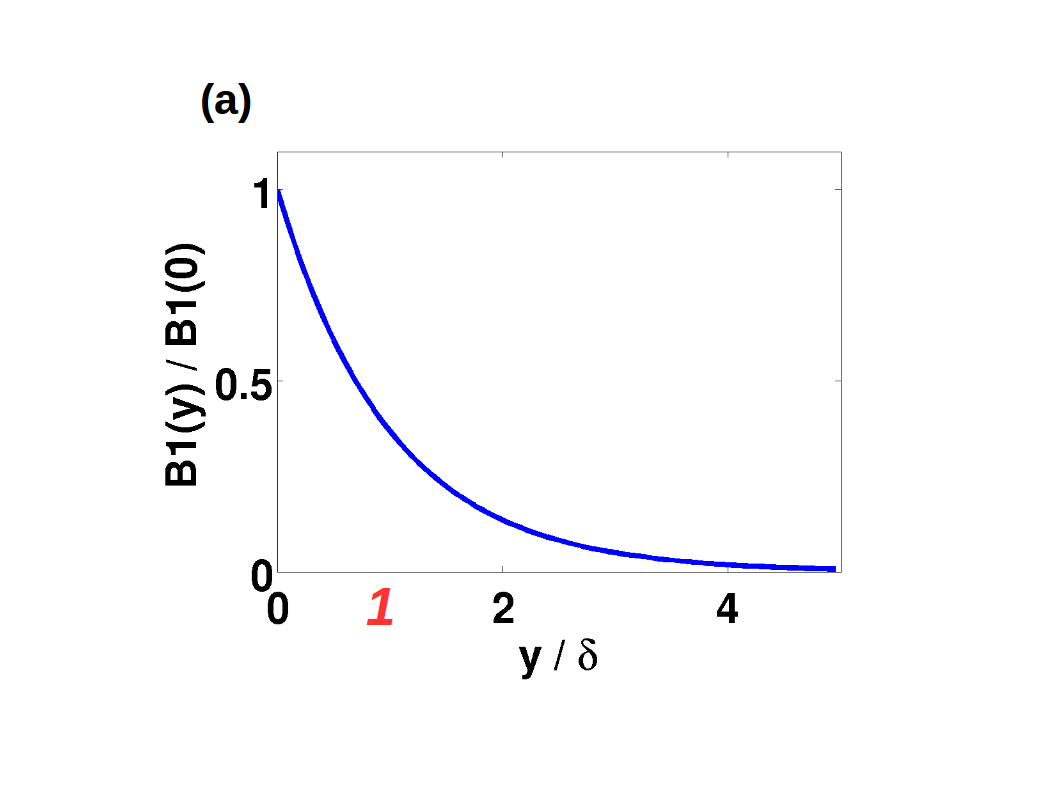}
\includegraphics[width=8cm]{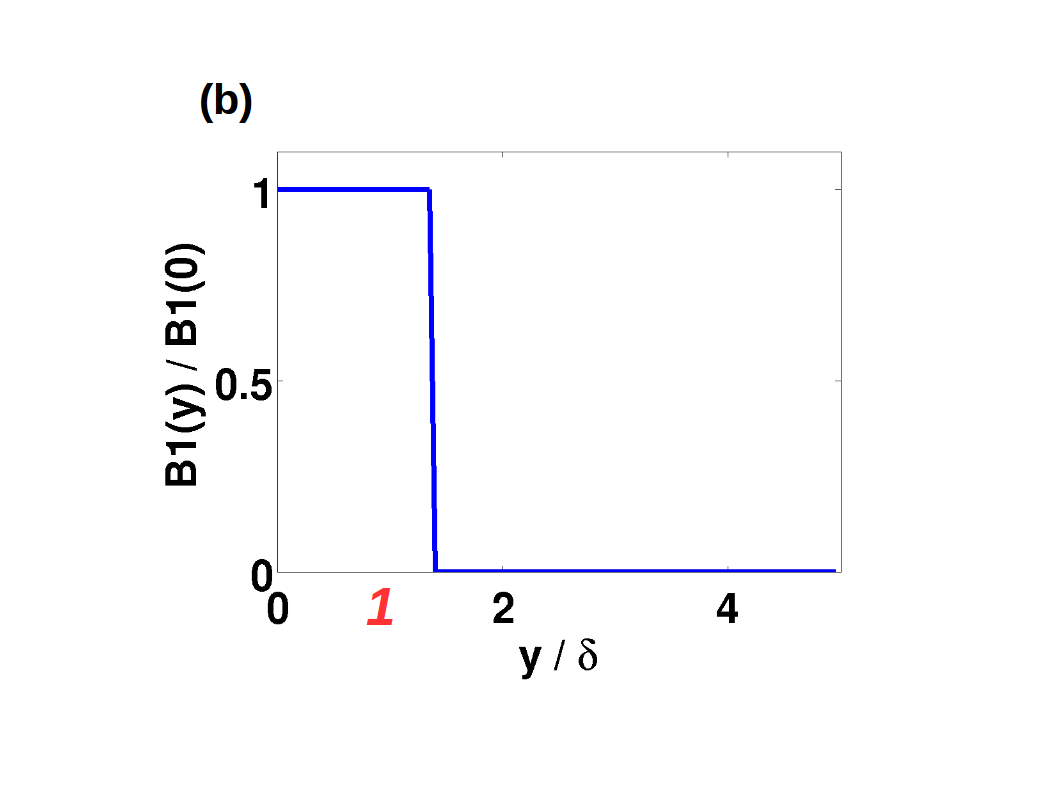}
\label{figSI0dltaEff}
\end{figure}

\begin{figure}
\caption{
{\bf Surprising regions of intensity in bulk metal MRI.}\\
\li7 2d MRI (sans slice selection) stack plots
of phantom P3 (Fig.\ref{fig0phntms} and
 section \ref{sxn0mthds}).
Vertical axis denotes intensity, resulting from the sum total of spin density
along non imaged dimension. 
\\
{\bf (a)} 2d MRI($xy$); same data as in blue, labeled P3, regions of
Fig.\ref{fig0xy}b.\\
{\bf (b)} 2d MRI($yz$).\\
In either image, there is no contribution from the $bc$ faces to the MR signal
since they are $\perp$ \Bone (section \ref{sxn0intrdxn}).
\\
In either image,
the high intensity {\em walls} are easily associated with the $ac$ faces.
(see Figs.\ref{fig0twoStrps}, \ref{fig0xy} and  section S\ref{sxnSI0thcknss}).\\
The intensity between the two $ac$ faces of the metal strip is evident in both
images.  Note the low intensity {\em plateau} and the {\em ridges} between the
walls, respectively in the MRI($xy$) and  MRI($yz$) images.
Regarding these novel regions of intensity, see section \ref{sxn0mri}.
See also Fig. \ref{fig0xy0yz0sim}.
}
\includegraphics[width=9cm]{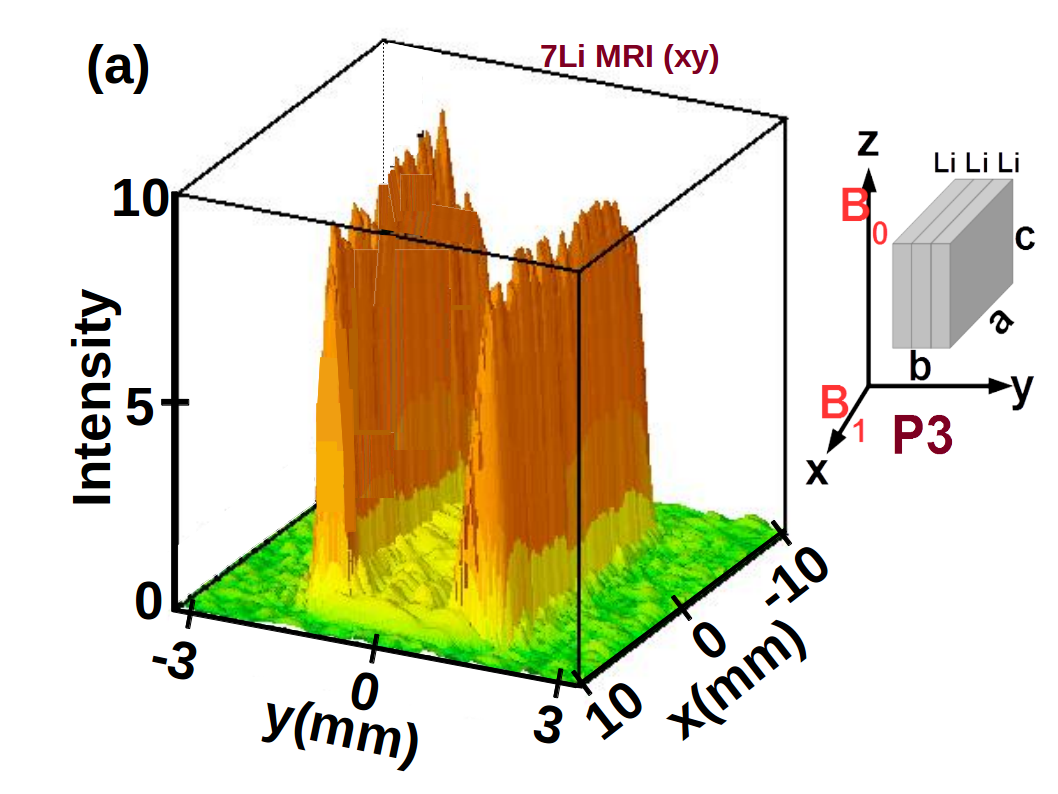}
\includegraphics[width=9cm]{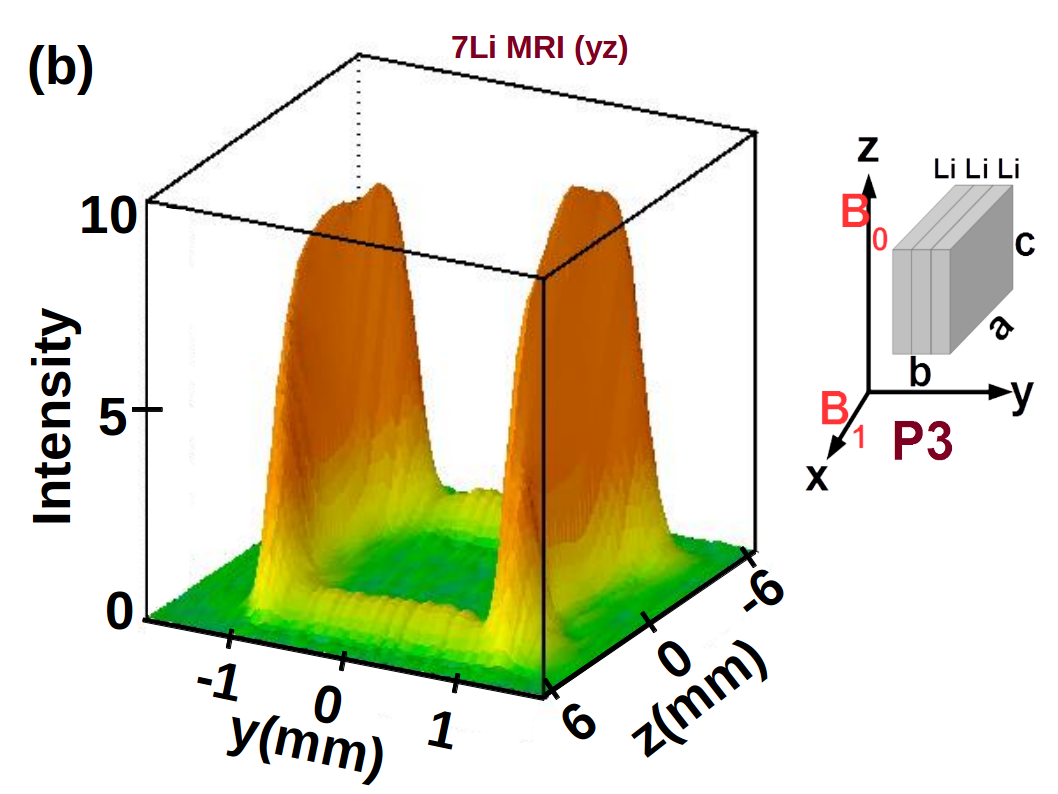}
\label{fig0xy0yz}
\end{figure}

\begin{figure}
\caption{
{\bf Peering at a virtual electromagnetic vacuum through an MRI {\em tunnel}.}\\
\li7 3d MRI($xyz$) of phantom P3 (Fig.\ref{fig0phntms}).
In addition to the already identified $ac$ faces in previous images
(Figs.\ref{fig0xy0yz},  \ref{fig0twoStrps} and \ref{fig0xy}),
the $ab$ faces are revealed for the first time,
accounting for the low intensity regions in Fig.\ref{fig0xy0yz}.
The $bc$ faces are conspicuous by absence, being $\perp$ \Bone.
The resulting {\em MRI tunnel} supplies a compelling visual of {\em peering} at
a virtual EM vacuum in the interior of a metallic conductor
(see section \ref{sxn0mri}).
}
\includegraphics[width=15cm]{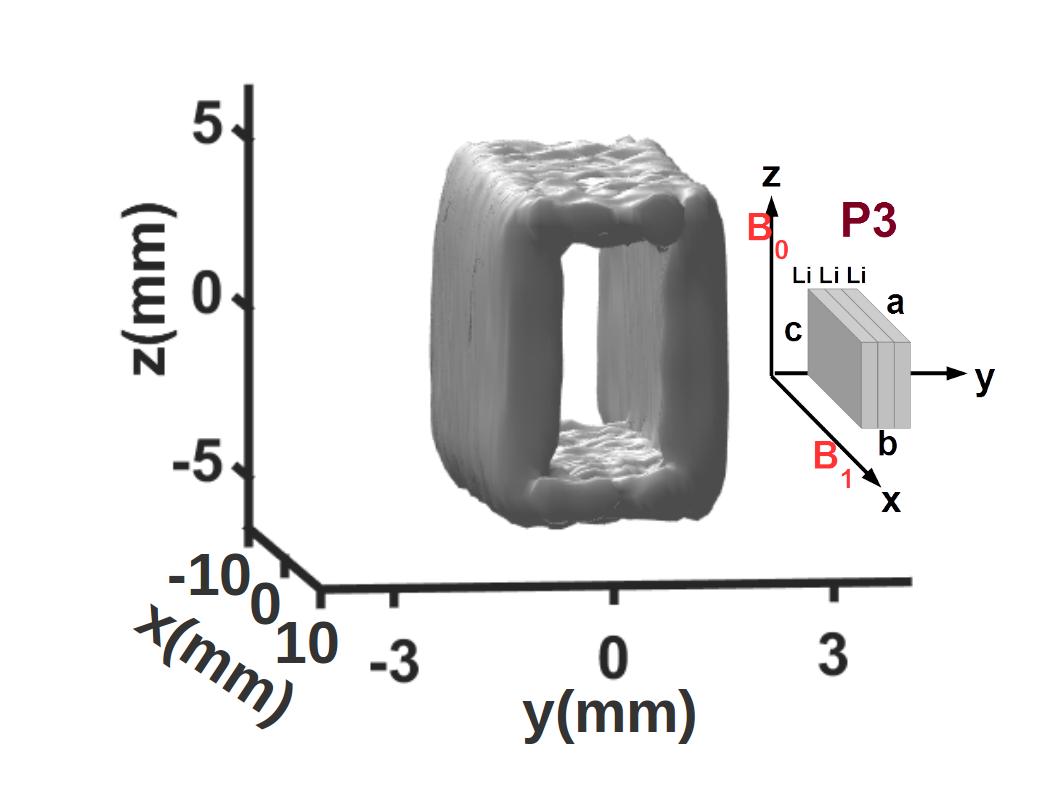}
\label{fig0xyz}
\end{figure}

\begin{figure}
\caption{
{\bf {\em Expected} relative intensities in bulk metal MRI.}\\
Illustration of {\em expected} relative intensities from different pairs of
faces in bulk metal \li7 2d MRI of phantom P3
as stack plots (intensity along vertical axis):
{\bf (a)} 2d MRI($xy$).
{\bf (b)} 2d MRI($yz$). \\
In either illustration, there is no contribution from the $bc$ faces to the MR
signal since they are $\perp$ \Bone (section \ref{sxn0intrdxn}).
\\
For MRI($xy$) in panel (a), the expected ratio of signal intensities from $ac$
and $ab$ faces ($S_{ac}/S_{ab} \approx 368 $) is in obvious disagreement with
experiment (Fig.\ref{fig0xy0yz}a). \\
For MRI($yz$) in panel (b), one would naively expect that 
$S_{ac}=S_{ab}$, which again is in striking departure from the experiment
(Fig.\ref{fig0xy0yz}b).
See section \ref{sxn0mri}.
}
\includegraphics[width=9cm]{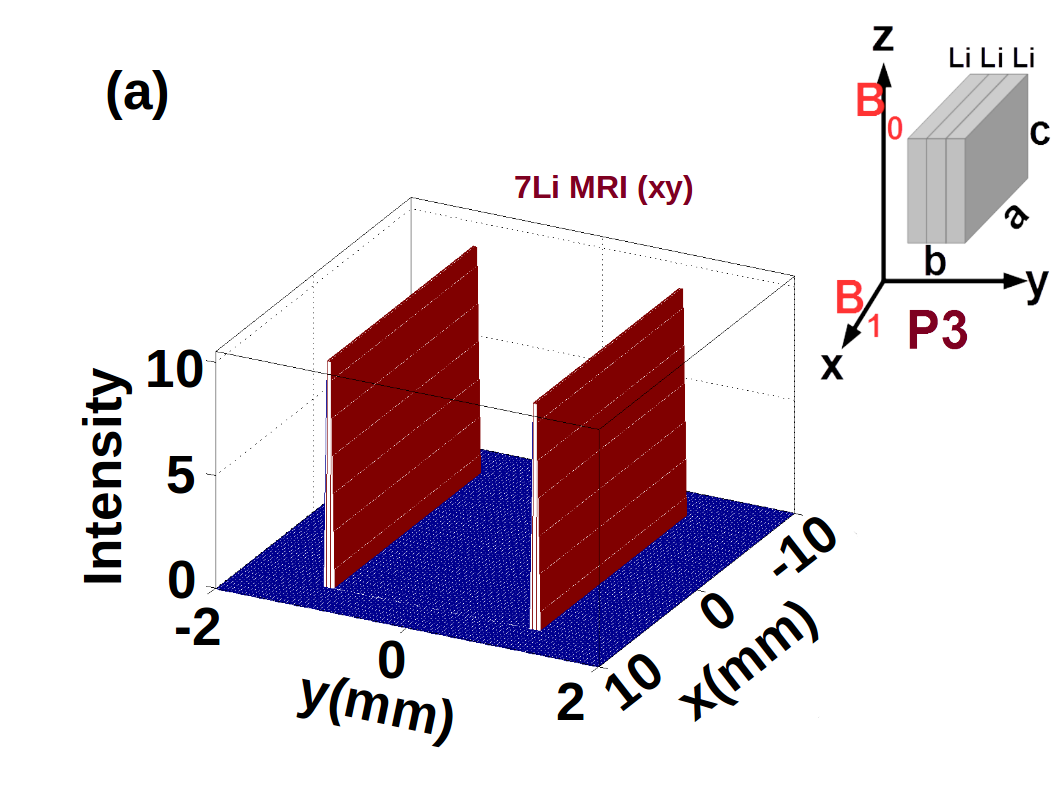}
\includegraphics[width=9cm]{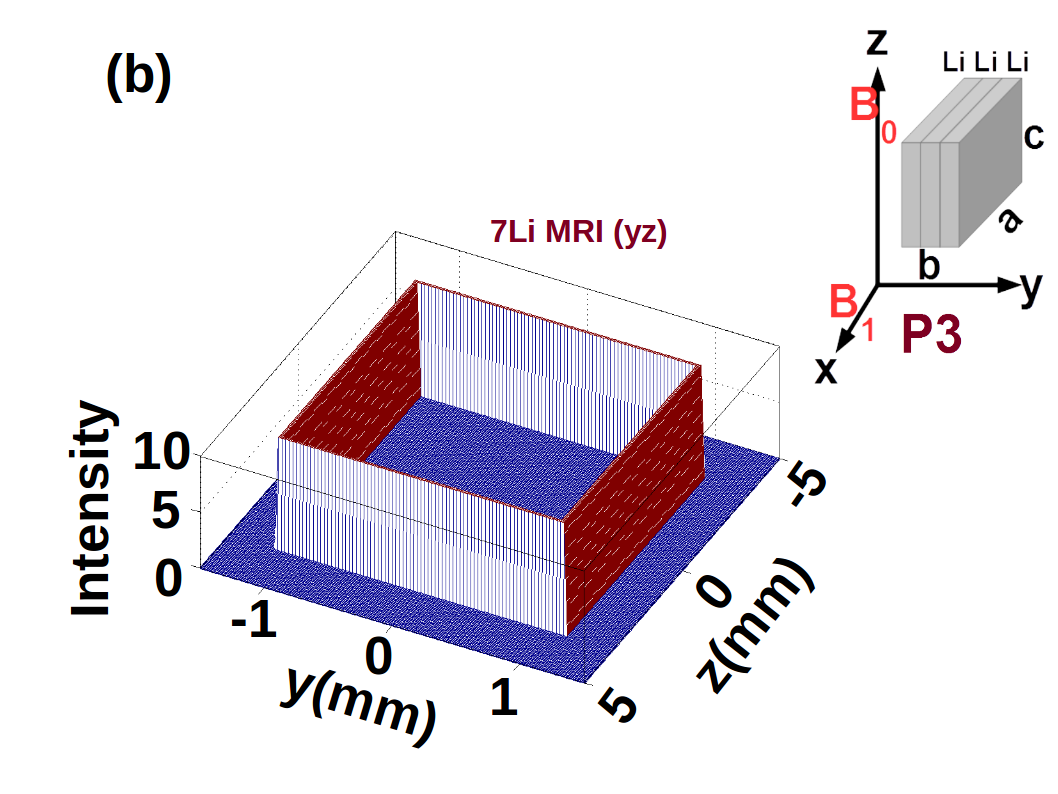}
\label{fig0xy0yz0sim}
\end{figure}

\begin{figure}
\caption{
{\bf Chemical shift imaging of a bulk metal strip.}\\
\li7 3d CSI($yz$) of phantom P3: chemical shift, $y$ and $z$ axes comprise the
three dimensions.
\\
Despite being composed of {\em identical} Li atoms, \li7 NMR spectrum (inset)
of phantom P3 exhibits two peaks, instead of the {\em expected} single peak
(section \ref{sxn0csi}).
Short and tall peaks are centered respectively at 
$\delta_1$= 256.4 and $\delta_2$= 266.3 ppm.
\\
In the CSI,
bands separated along $z$, occur at CS $\delta_1$. The
bands separated along $y$, occur at CS $\delta_2$. \\
In conjunction with the 3d MRI (Fig.\ref{fig0xyz}) and P3 schematics
(Fig.\ref{fig0phntms}),
the pair of bands at $\delta_1$ are assigned to $ab$ faces,
and pair of bands at $\delta_2$ are assigned to $ac$ faces
(thus completing the assignment of both the 2d CSI($y$) in Fig.\ref{figSI0csi}, 
and the NMR spectrum itself).
}
\includegraphics[width=12cm]{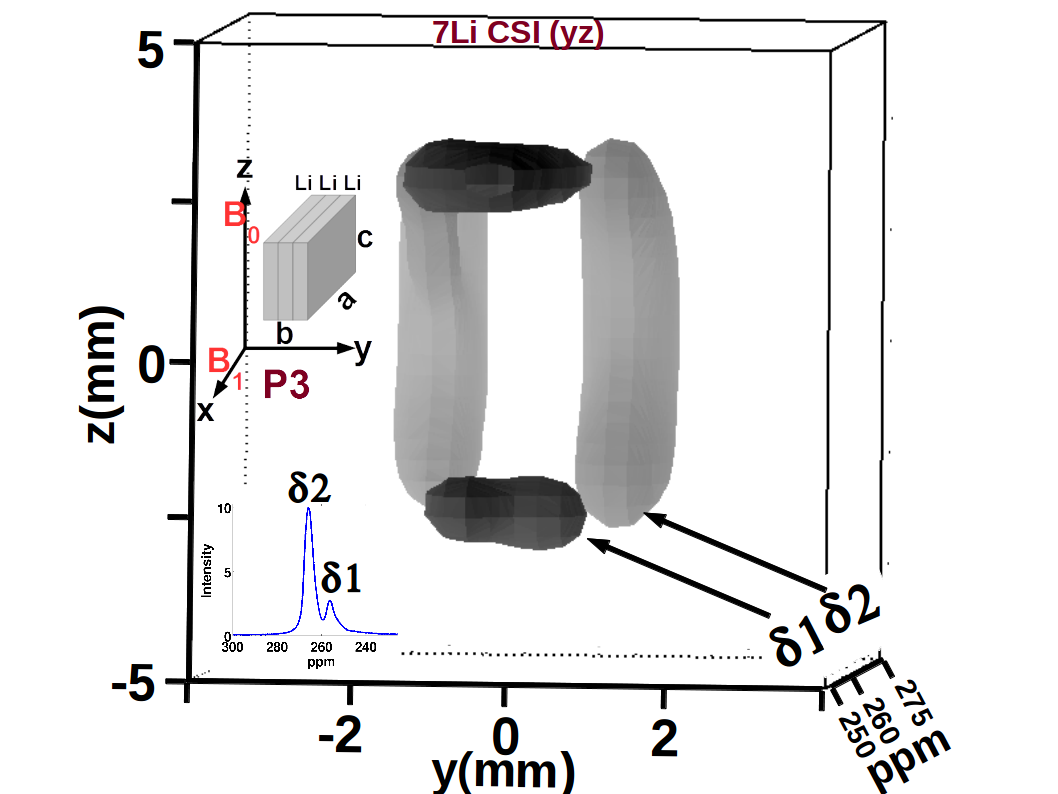}
\label{fig0csi}
\end{figure}

\begin{figure}
\caption{
{\bf {\em Predicted} relative intensities in bulk metal MRI.}\\
Illustration of {\em predicted} (from the derived formulae) relative intensities
from different pairs of faces in bulk metal \li7 2d MRI of phantom P3
as stack plots (intensity along vertical axis). \\
{\bf (a)} 2d MRI($xy$) based on Eq.(\ref{eq:eq0ratioxy}): $S_{ac}/S_{ab}=14$. \\
{\bf (b)} 2d MRI($yz$) based on Eq.(\ref{eq:eq0ratioyz}): $S_{ac}/S_{ab}=28$. \\
In either illustration, there is no contribution from the $bc$ faces to the MR
signal since they are $\perp$ \Bone (section \ref{sxn0intrdxn}).
\\
Compare and contrast with the corresponding experimental images in
Fig.\ref{fig0xy0yz}, and the illustration of {\em expected} images in
Fig.\ref{fig0xy0yz0sim}.
See sections \ref{sxn0mri} and \ref{sxn0formulae}.
}
\includegraphics[width=9cm]{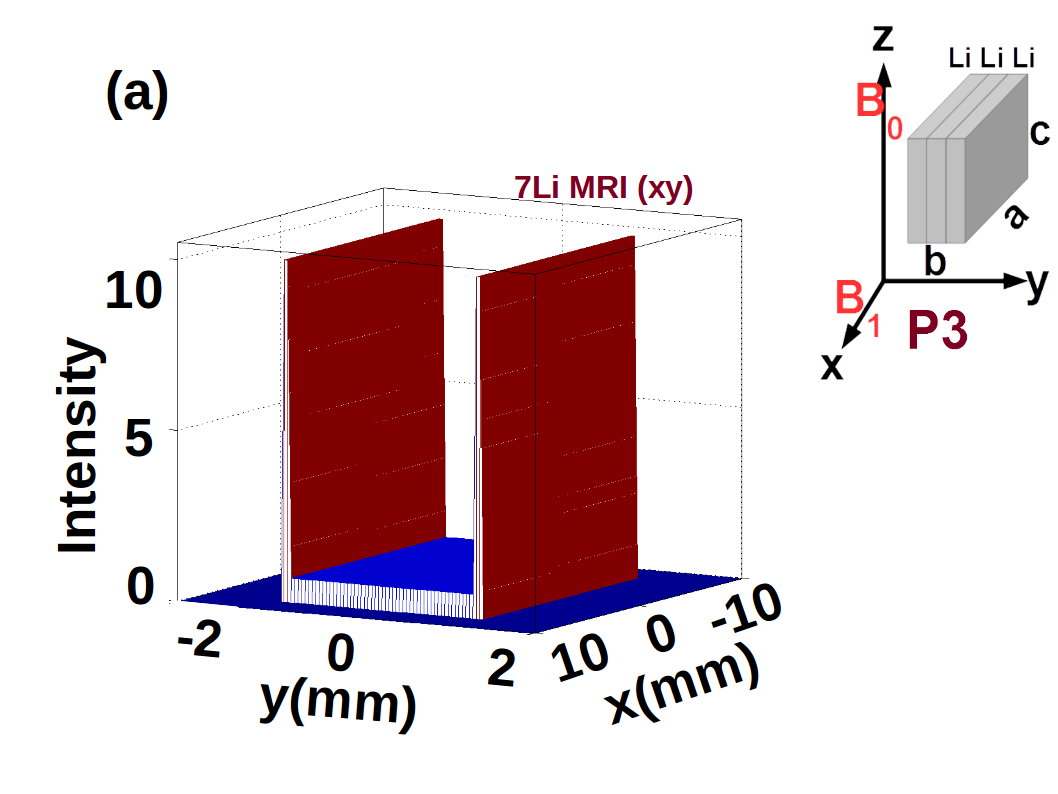}
\includegraphics[width=9cm]{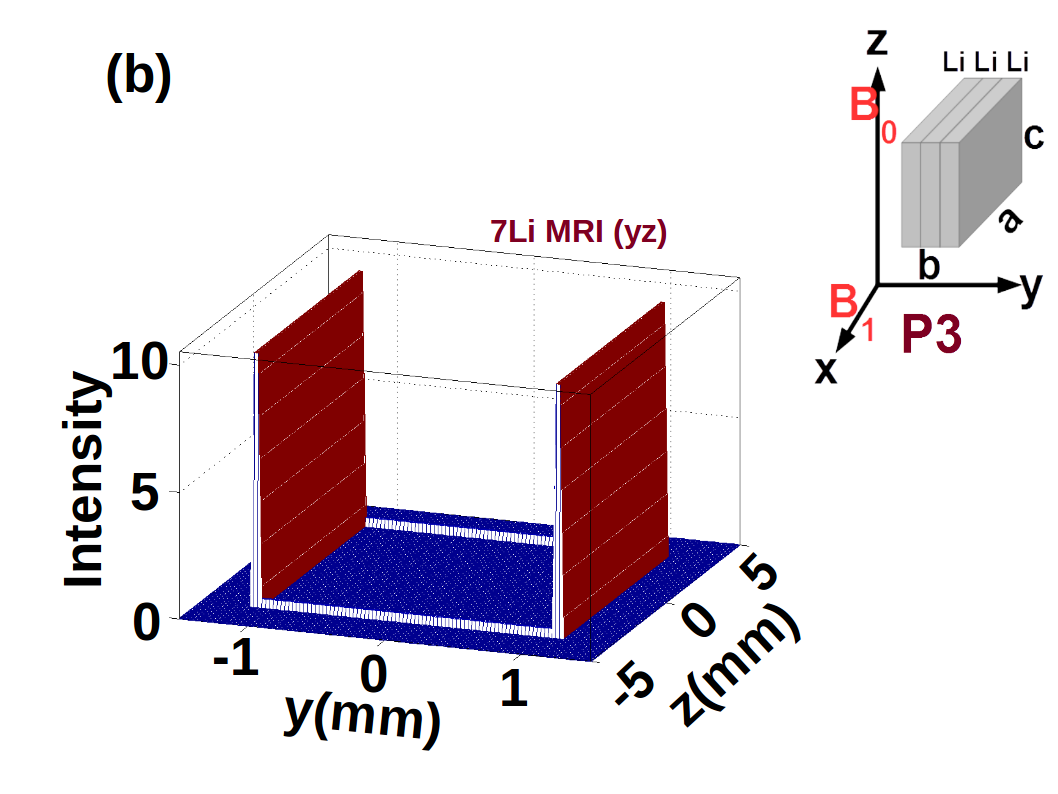}
\label{fig0xy0yz0drvd}
\end{figure}

\clearpage

\title{
{\bf SUPPORTING INFORMATION} \\
  for\\
{\bf Visualizing electromagnetic vacuum by MRI}
}

\authors{ 
Chandrika S Chandrashekar$^1$,
Annadanesh Shellikeri$^2$,
S Chandrashekar$^{3*}$,
Erika A Taylor$^4$, Deanne M Taylor$^{5,6,7}$
}


\affil{
1. Lincoln High School (class of 2018), 3838 Trojan Trail, Tallahassee,
Florida, 32311, USA \\
2. Aeropropulsion, Mechatronics and Energy Center, Florida State University,
2003 Levy Ave., Tallahassee, FL 32310, USA \\
3. National High Magnetic Field Laboratory (NHMFL) and Florida State University,
1800 E. Paul Dirac Drive, Tallahassee, Florida, 32310, USA \\
4. Department of Chemistry, Wesleyan University,52 Lawn Ave., Hall-Atwater Labs,
Middletown, Connecticut 06459, USA \\
5. Department of Pediatrics, Perelman School of Medicine,
University of Pennsylvania, Philadelphia, Pennsylvania, 19104, USA \\
6. Department of Biomedical and Health Informatics,
The Children's Hospital of Philadelphia, Philadelphia, PA 19041, USA\\
7. Department of Genetics, Rutgers University, Piscataway, NJ 08854, USA\\
{\bf ($^*$ Corresponding author)} \\
start:  {\bf 20150601}
posted: {\bf 20160912}
rvsn1:  {\bf 20161124}
rvsn2:  {\bf 20161208}
}

\begin{article}

\renewcommand{\theequation}{S\arabic{equation}}
\renewcommand{\thesection}{S\arabic{section}}
\renewcommand{\thesubsection}{\thesection.\arabic{subsection}}



\section{ Effective subsurface depth, $\delta_{\text{eff}}$}
\label{sxnSI0subSurface}
The r.f. EM fields, upon encountering bulk metal, face the
phenomenon of {\em skin effect}.  Of particular interest to this work, is the
fact that the magnitude of the r.f. magnetic field (\Bone) decays exponentially 
inside the metal according to
\cite{jackson,si0griffiths,si0ulaby}
\begin{equation}
 B_1(y)= B_1(0) e^{-y/\delta}
\label{eq:eq0b1exp}
\end{equation}
at a depth $y$ beneath the surface as shown in Fig.\ref{figSI0dltaEff}a.
The characteristic length $\delta$ (the skin depth), is determined by:
\begin{equation}
 \delta= \sqrt{\frac{2}{\mu \sigma \omega_{\text{r.f.} } } }
\label{eq:eq0sknDpth}
\end{equation}
where, $\sigma$ is the conductivity of the conductor (metal),
$\mu$, its magnetic permeability, with, 
\begin{equation}
 \mu= \mu_0\ \mu_r;\ \ \ \  \sigma= 1/\rho;\ \ \ \
 \omega_{\text{r.f.} }= 2\pi \nu_{\text{r.f.} }
\label{eq:eq0wRF}
\end{equation}
where,
$\rho$\ is the resistivity of the metal,
$\mu_r$ is the relative magnetic permeability of the metal {\em relative} to
$\mu_0$, the free space permittivity, and
$\nu_{\text{r.f.} }$ is the (radio) frequency of the applied field.

$\mu_0$= 4$\pi$\ 10$^{-2} m kg A^{-2} s^{-2}$  \cite{si0CRC1997handbook}. 
For Li, $\rho$= 92.8 $n\Omega\ m,\  \mu_r$=1.4 \cite{si0CRC1997handbook}.
$\nu_{\text{r.f.}}$= 350 $MHz$, is the frequency of the applied r.f. field,
set to the resonance (Larmor) frequency for \li7 in a magnetic field of strength
21 $T$.  When used in Eq.(\ref{eq:eq0sknDpth}), these values yield
$\delta \approx 6.9 \mu m$.
\\

Consider a face $\parallel$ \Bone. Using Eq.(\ref{eq:eq0b1exp}),
the subsurface beneath this face, per unit surface area, contributes to the
signal according to \cite{rangeet},
\begin{equation}
 S \propto \int_0^{\infty} dy\ \sin (\theta_0\ e^{-y/ \delta} )
\label{eq:eq0sgnlExp}
\end{equation}
where,
\begin{equation}
  \theta_0 \equiv \theta(0)= \gamma B_1(0)\ \tau 
\label{eq:eq0flpAngl}
\end{equation}
is the r.f. flip angle \cite{callaghan,haacke,abragam,si0ernst} at the 
metal surface ($y$=0) \cite{rangeet},
$\gamma$, the gyromagnetic ratio for \li7 nucleus,
$\tau$, the duration of the applied r.f. pulse.
Eq.(\ref{eq:eq0sgnlExp}) may be readily recast as:
\begin{equation}
 S \propto \delta \int_0^{\theta_0} dy\ \text{sinc}(y)
   \equiv \delta\ Si(\theta_0)
\label{eq:eq1sgnlExp}
\end{equation}
where, sinc$(y)= \sin y/y$.
The well known "Sine" integral $Si$ can be evaluated by numerical
integration (for e.g., using the function "sinint" of popular mathematical
software MatLab).

In particular, for flip angle $\theta_0$=\piByTwo\ \cite{rangeet},
using Eq.(\ref{eq:eq1sgnlExp}), the signal contribution, from the subsurface
per unit surface area is
\begin{equation}
S \propto 1.3708\ \delta \equiv \delta_{\text{eff}}
\label{eq:eq0dltaEff}
\end{equation}
where, $\delta_{\text{eff}}$ is the {\em effective} subsurface depth
contributing to the signal,
in the {\em absence} of exponential decay of \Bone (Fig.\ref{figSI0dltaEff}b).
We can generalize the proportionality above to an arbitrary elemental surface
area ($\perp y$):
\begin{equation}
S \propto  dx\ dz\ \delta_{\text{eff}}
\label{eq:eq1dltaEff}
\end{equation}
For our case, from Eq.(\ref{eq:eq0dltaEff}), $\dltEff \approx 9.49 \mu m$
(Fig.\ref{figSI0dltaEff}b).

\section{Metal strip thickness from MRI}
\label{sxnSI0thcknss}
\subsection{Phantom P0 with 2 Li strips}
\label{subsxnSI2strps}
The now familiar \cite{chandrashekar,ilott,romanenko} bulk metal two
dimensional
(2d) MRI($xy$), for phantom P0 (Fig.\ref{fig0phntms})
is shown in Fig.\ref{fig0twoStrps}a.
The  extent of image intensity bands along the direction of separation ($y$),
arises from each ($b$=0.75 $mm$ thick) conducting metal strip,
and is mainly determined by the image resolution (0.5 $mm$) along $y$
(Methods section \ref{subsxn0mthdsMri}).
It is difficult to deduce the strip thickness accurately
(each band yielding 1.8 $mm$ for each strip).

The situation is quite different at {\em twice} the resolution
along $y$, as shown in Fig.\ref{fig0twoStrps}b.
For the higher resolution image, each conducting metal strip gives rise to two
bands, from the $ac$ faces
$\parallel$ \Bone, since the magnitudes of strip thickness, skin depth and
resolution collude together to resolve the faces of each strip
\cite{ilott,romanenko}.
The strip thickness is thrice the resolution and is far greater (79 x) than
$\dltEff\ (=9.49 \mu m$ for our case, as shown in section
S\ref{sxnSI0subSurface}; 
the two effective subsurfaces beneath the $ac$ faces are separated by 731
$\mu m$).
The intra strip band separation yields a measured thickness of 
0.80 and 0.79 $mm$ respectively for top and bottom strips (a ratio of 0.99).
By contrast, for the image in panel (a), the signals from the two faces of a
given strip coalesce into a single band due to inadequate resolution.

\subsection{Phantoms P1 and P3}
\label{subsxnSI0p1p3}
Fig.\ref{fig0xy} displays an overlay of \li7
1d MRI($y$)  in panel (a), and 2d MRI($xy$) in panel (b),
from phantoms P1 and P3.
The nominal resolutions along $y$, for 1d MRI($y$) and 2d MRI($xy$), are
respectively  0.0357 and 0.25 $mm$.
Each phantom gives rise to two bands emanating from its {\em pair} of $ac$
faces.
The separaton (along $y$) amongst the intra-strip bands yields the strip
thickness.

It is evident that the super strip of phantom P3, composed of 3 Li strips
pressed together behaves as a single strip.
Mere visual inspection of the images, by virtue of the equal spacing
between the 4 bands along $y$, suggests that P3 is thrice the thickness of 
P1.

The thickness for P3 and P1 are respectively,
2.163 and 0.73 $mm$ from the 1d images, and 
2.51  and 0.83 $mm$ from the 2d images,
verifying that P3 is $\approx$ 3x thicker than P1.
\\

Thus, the drawback of limited r.f. penetration and attenuated signal
in bulk metals, due to skin-effect, can be turned into an opportunity
to undertake noninvasive thickness measurements \cite{romanenko}.

\section{{\em Peculiar} intensity ratios in bulk metal MRI and CSI}
\label{sxnSI0mri2dNtnst}
In section \ref{sxn0formulae} we proposed that gradient switching during MRI
experiments could be responsible for the observed intensity differences of MRI
and CSI signals from $ab$ and $ac$ faces of phantom P3.  Here we examine this
in detail.

Eddy currents \cite{jackson,si0griffiths,si0saslow}, produced by the gradient
switching could adversely affect the MRI signal
\cite{callaghan,haacke,romanenko} from the $ab$ and $ac$ faces
by differing amounts.  The transient magnetic fields ($\parallel$ \B0)
produced during the gradient switching, induce eddy currents in {\em closed}
loops on a $\perp$ surface, which in turn produce opposing magnetic fields
according to Lenz's Law \cite{jackson,si0griffiths,si0ulaby}.
It is the induced, instead of the instigating, magnetic fields that are of
interest, since the eddy currents can persist long after the gradient switching.
The induced magnetic fields can alter the precession frequencies of the
spins in the transverse ($\perp$ to \B0) plane leading to phase variations,
diminished signal, and image distortions.
The $ac$ face is $\parallel$ to the transient magnetic fields of the gradient,
and hence unable to support a closed current loop to exist on its surface
\cite{si0brey}, while the $ab$ face can.
Thus, in general, the $ac$ and $ab$ faces can have different signal intensities.
\\

To test this hypothesis, we compared
the \li7 2d MRI($yz$) in vertical   ($ac \parallel \B0$, $ab \perp \B0$), and
the MRI($zy$) in horizontal ($ab \parallel \B0$, $ac \perp \B0$)
orientations, shown in Fig.\ref{fig0p3hrzntl}.

In the vertical orientation, signal intensity from the
$ac$ face should not be affected by the eddy currents while that from the $ab$
face would be.  In the horizontal orientation, it would be {\em vice versa}.

But, the MRI($yz$) in vertical orientation and MRI($zy$) in horizontal
orientation (Fig.\ref{fig0p3hrzntl}), show this not to be the case.
They were acquired under completely equivalent conditions
(Methods section \ref{subsxn0mthdsMri}),
resulting in a virtual dead heat regarding the relative 
intensities from the $ab$ and $ac$ faces in these two orientations.\\ 
(For MRI($yz$) the ratio of signal intensities from $ac$ and $ab$ faces is 10
(section \ref{sxn0mri}), while for MRI($zy$) it is 10.1.
Also, the corresponding NMR spectra, shown as insets, confirm our
assignments 
of downfield and upfield peaks respectively to faces $\parallel$ and
$\perp$ to \B0 as described in section \ref{sxn0csi},
and are consistent with the reported experiments and simulations
\cite{rangeet,ilott}.)
\\
By the same token, the eddy currents produced by the r.f. \Bone \cite{si0balcom}
would not affect $ab$ and $ac$ faces, which are always $\parallel \Bone$ for our
case
(indirectly corroborated by prior studies
\cite{rangeet,chandrashekar,ilott,hjc2015,lina,hjc},
that correctly accounted for NMR signal from metal strips).
On the other hand, face $bc \perp \Bone$, and the resultant eddy current
annihilates \Bone \cite{jackson,si0griffiths} and MR signal for this face
\cite{chandrashekar,ilott}, as borne out time and again by the MRI
and CSI images throughout this manuscript.

It should be noted that modern gradient systems have built in active
shielding technology to largely suppress the formation of eddy currents to
mitigate the deleterious effects on the MR signal
\cite{callaghan,haacke,si0brey1987}, rendering 
the vertical and horizontal orientations (Fig.\ref{fig0p3hrzntl}) equivalent
in this regard. \\

\section{\Bone\ inhomogeneity.}
One might ask if the intenstiy differences from $ab$ and $ac$ subsurfaces
in the MRI and CSI images are due to differences in \Bone\ amplitude
(\Bone\ {\em inhomogeneity}) at the two orthogonal faces.

Consider the 2d MRI($xy$) images as a function of resolution in
Fig.\ref{figSI0td20td40td80}.
Let $B_1^{xy}$ and $B_1^{xz}$ denote the $B_1$ amplitudes respectively at the
$ab$ and $ac$ faces, and $f(B_1^{xy})$ and $f(B_1^{xz})$ the corresponding
functions affecting the signal intensities from these faces.
Assuming equal elemental volumes (voxels) underneath these faces, the intensity
ratios for panels-a,b,c, would respectively be,
\begin{equation}
 \frac{S_{ac}}{S_{ab}}= \left\{
  \frac{\Delta x\ (\Delta y/q)\ c}
       {\Delta x\ (\Delta y/q)\ c} \right\}
  \frac{f(B_1^{xz})} {f(B_1^{xy})}
 =\frac{f(B_1^{xz})} {f(B_1^{xy})};\ \ \ \  q= 1, 2, 4
\label{eq:eq1ratioxy}
\end{equation}
since resolution along $y$ increases respectively by factors of 1, 2 and 4, and
$c$ is length of the metal strip along $z$ (the spatially unresolved dimension).
Eq.(\ref{eq:eq1ratioxy}) predicts that the ratios are independent of resolutions
along $x,y,z$, in contradiction to the experiment.

On the other hand, the existence of \Bone\ differences at the $ab$ and $ac$
faces should have resulted in different $S_{ac}/S_{ab}$ for the two images in
the {\em equivalence} experiment in Fig.\ref{fig0p3hrzntl}.
For the 2d MRI($yz$) in vertical orientation in Fig.\ref{fig0p3hrzntl}a, the
intensity ratio would be,
\begin{equation}
 \frac{S_{ac}}{S_{ab}}= \left\{
  \frac{a\ \Delta y\ \Delta z} {a\ \Delta y\ \Delta z} \right\}
  \frac{f(B_1^{xz})} {f(B_1^{xy})} =\frac{f(B_1^{xz})} {f(B_1^{xy})}
\label{eq:eq2ratioyz}
\end{equation}
assuming again equal elemental volumes underneath these faces.
Also, $a$ is length of the metal strip along $x$
(the spatially unresolved dimension).
Similarly, for the 2d MRI($yz$) in horizontal orientation in
Fig.\ref{fig0p3hrzntl}b, the intensity ratio would be:
\begin{equation}
 \frac{S_{ac}}{S_{ab}}= \frac{f(B_1^{xy})} {f(B_1^{xz})}
\label{eq:eq2ratiozy}
\end{equation}
Thus, the prediction based on \Bone\ inhomogeneity, via
Eqs.(\ref{eq:eq2ratioyz}) and (\ref{eq:eq2ratiozy}),
is again contrary to the observation, since the ratio of the intensity ratios
would be $[f(B_1^{xz}) / f(B_1^{xy}) ]^2$ (=0.98, experimentally).
This implies, practically $f(B_1^{xz}) = f(B_1^{xy})$.

We can also examine if the \Bone\ inhomogeneity could {\em partially} account
for the descrepancy between observed and derived $S_{ac}/S_{ab}$ values. 
Consider the 2d MRI($yz$) in vertical orientation shown in
Fig.\ref{fig0p3hrzntl}a.
The differing $B_1$ amplitudes $B_1^{xy}$ and $B_1^{xz}$ respectively at 
the $ab$ and $ac$ faces, should result in corresponding effective subsurface
depths $\dltEff^{xy}$ and $\dltEff^{xz}$ (see Eqs.
(\ref{eq:eq0b1exp})-(\ref{eq:eq1dltaEff})), replacing Eq.(\ref{eq:eq0ratioyz})
by
\begin{equation}
 \frac{S_{ac}}{S_{ab}}= \frac{\Delta z}{\Delta y}\
   \frac{\dltEff^{xz}}{\dltEff^{xy}}
\label{eq:eq1ratioyz}
\end{equation}
Similarly, for the 2d MRI($zy$) in horizontal orientation shown in
Fig.\ref{fig0p3hrzntl}b, Eq.(\ref{eq:eq0ratiozy}) is replaced by
\begin{equation}
 \frac{S_{ac}}{S_{ab}}= \frac{\Delta y}{\Delta z}\
   \frac{\dltEff^{xy}}{\dltEff^{xz}}
\label{eq:eq1ratiozy}
\end{equation}
But, the experimental setup guarantees that the resolution ratios in both
Eq.(\ref{eq:eq1ratioyz}) and Eq.(\ref{eq:eq1ratiozy}) is 28.  Hence, the
ratio of the intensity ratios should be
$[\dltEff^{xz} / \dltEff^{xy}]^2$ (=0.98, experimentally),
leading to the conclusion that \Bone\ inhomogeneity does not play a prominent
role in this study.
(By the numbers:
Using the experimental intensity ratio in Eq.(\ref{eq:eq1ratioyz}),  
$\dltEff^{xz} / \dltEff^{xy}= 10/28$. Using this value in
Eq.(\ref{eq:eq1ratiozy}), we see that the intensity ratio should have been
78.4, in contradiction with the experimental value of 10.1, supporting the claim
that \Bone\ inhomogeneity does not have an appreciable contribution.) \\

The {\bf strongest support yet for this view comes from prior studies}
\cite{rangeet,chandrashekar,ilott,hjc2015,lina,hjc}, relying on skin depth
arguments, that correctly accounted for the MR signal from metal strips,
{\em without} invoking \Bone\ inhomogeneity.

Consider for example the formula for bulk metal signal, given by
Supplementary Information Eq.(4) of Reference \cite{rangeet}:
\begin{equation}
 S_{bulk}= 2\ S_0\  \dltEff\ (ab + bc + ca)
\label{eq:eq0sgnlBulk}
\end{equation}
where, \dltEff\ is given by Eq.(\ref{eq:eq0dltaEff}).
If \Bone\ amplitudes were different for different types of faces of the metal
strip, Eq.(\ref{eq:eq0sgnlBulk}) would be replaced by:
\begin{equation}
 S_{bulk}= 2\ S_0\  (ab\ \dltEff^{ab}  + bc\ \dltEff^{bc}  + ca\ \dltEff^{ac} )
\label{eq:eq1sgnlBulk}
\end{equation}
accounting for differing effective subsurface depths resulting from differing
\Bone\ amplitudes at the metal strip surfaces.
Since, Eq.(\ref{eq:eq0sgnlBulk}) has been established to be correct
\cite{rangeet}, we can conclude from Eq.(\ref{eq:eq1sgnlBulk}) that:
\begin{equation}
\dltEff^{ab}=\dltEff^{bc}=\dltEff^{ac} \equiv \dltEff
\label{eq:eq0dltEffsEqul}
\end{equation}
Thus, it is unlikely that \Bone\ inhomogeneity accounts for the intensity
differences of signals associated with diferent faces of a metal strip. \\

Present study extends the established approach to MRI and CSI of bulk metals,
by following the same principles.
Identifying other mechanisms at play that could account for the remaining
discrepancy
between observed and derived intensity ratios, warrants further future research.
Nevertheless, the bulk metal MRI and CSI intensity patterns derived from our
formulae (based on {\em effective elemental volumes}) qualitatively and
quantitatively bear closer resemblance to experiment, than what is
conventionally expected (based on skin depth consideratons alone or traditional 
specifications of the voxel= $\Delta x \Delta y \Delta z$).

\end{article}

\clearpage

\setcounter{figure}{0}
\renewcommand{\thefigure}{S\arabic{figure}}

\begin{figure}
\caption{
{\bf Schematics of MRI {\em  phantoms.}}\\
Phantoms comprising of Li strips (of dimensions $a \times b \times c$), derived 
from a 0.75 $mm$ thick stock Li strip.\\
Phantom {\bf P0}: Pair of Li strips separated by a teflon strip;
For each Li strip,
$a \times b \times c=$ 20 x 0.75 x 7 $mm$$^3$.
\\
Phantom {\bf P1}: Single Li strip.
$a \times b \times c=$ 15 x 0.75 x 7 $mm$$^3$.
\\
Phantom {\bf P3}: Three  Li strips pressed together to yield a single
composite super strip.
$a \times b \times c=$ 15 x 2.25 x 7 $mm$$^3$.
\\
The static (main) magnetic field \B0 specifies the $z$ direction.
Also, $x,y,z$ denote the MRI gradient (imaging) directions.

The setup of phantom, r.f. coil and the gradient assembly guarantees that
$x \parallel a \parallel \Bone$, with a rotational degree of freedom about the
$x$-axis, to reorient the phantom.  See sections \ref{sxn0intrdxn} and
\ref{subsxn0mthdsPhntms}.
}
\includegraphics[width=8cm]{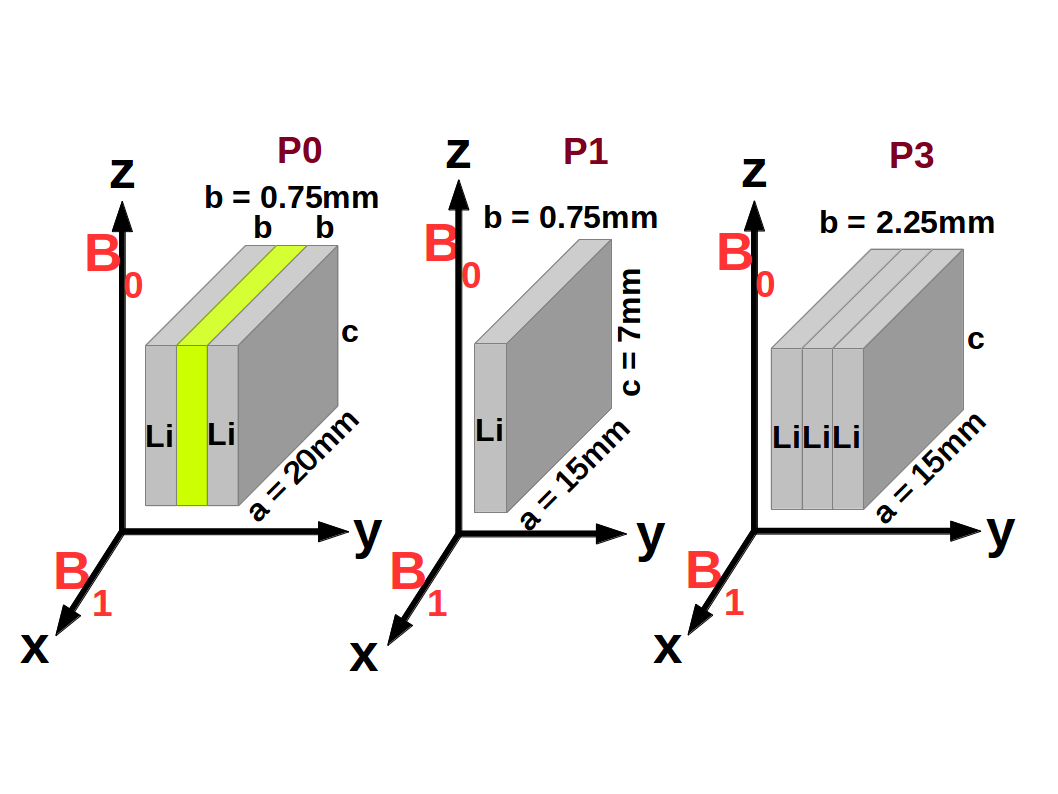}
\label{fig0phntms}
\end{figure}

\begin{figure}
\caption{
{\bf Resolution and strip thickness in bulk metal MRI.}\\
\li7 2d MRI(xy) from phantom P0 comprising of two Li
strips (separated by a teflon strip) of identical thickness, at
({\bf a}) 0.5 $mm$  ({\bf b}) 0.25 $mm$ resolution, along $y$.
\\
For the lower resolution image in panel (a), the  extent of image intensity
bands along $y$ is mainly determined by the image resolution.
It is difficult to infer the strip thickness with any confidence.\\
On the other hand, for the higher resolution image in panel (b),
each conducting strip gives rise to two bands, from the $ac$ faces
$\parallel \Bone$; the relative magnitudes of strip thickness, skin-depth and
resolution conspire and combine to resolve the faces of each strip.
The intra strip band separation yields the thickness of that strip.\\
See section S\ref{subsxnSI2strps}.
}
\includegraphics[width=14cm]{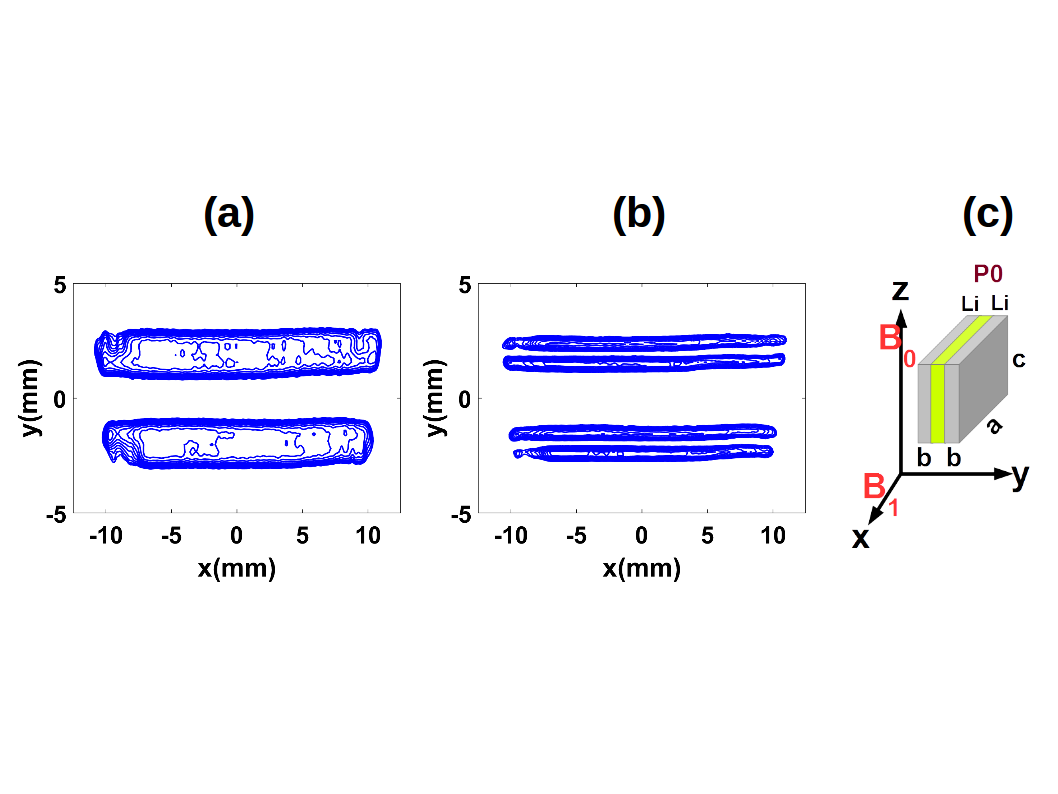}
\label{fig0twoStrps}
\end{figure}

\begin{figure}
\caption{
{\bf Comparing thickness of strips from bulk metal MRI.}\\
Superpositon of \li7 MRI from phantoms P1 and P3.
({\bf a}) 1d MRI(y).  The resolution along $y$, is 0.0357 $mm$.
({\bf b}) 2d MRI(xy).  The resolution along $y$, is 0.250 $mm$. \\
$x,y,z$ are the imaging directions (Fig.\ref{fig0phntms}).
Each phantom gives rise to two bands emanating from its {\em pair} of $ac$
faces.
Thickness of a given strip is given by the separation, along $y$, between bands
arising from it.
Mere visual inspection of the panels, by virtue of the equal spacing
between the 4 bands along y, suggests that P3 is thrice the thickness of P1. 
See section S\ref{subsxnSI0p1p3}.
}
\includegraphics[width=10cm]{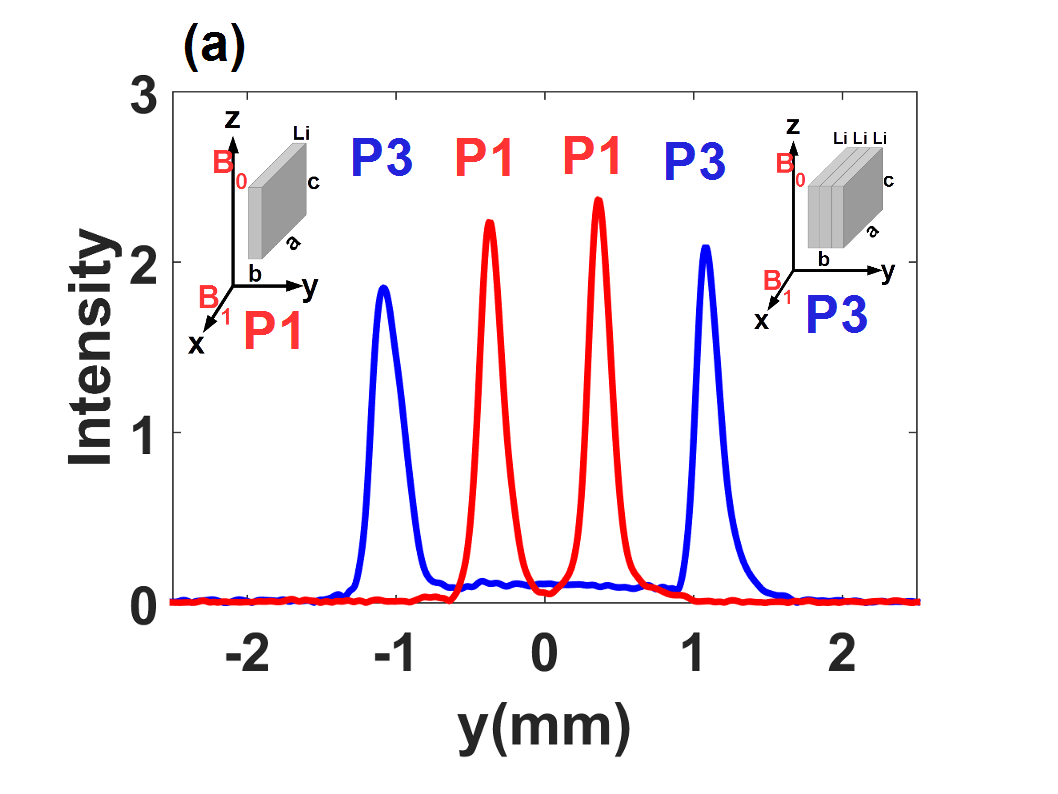}
\includegraphics[width=10cm]{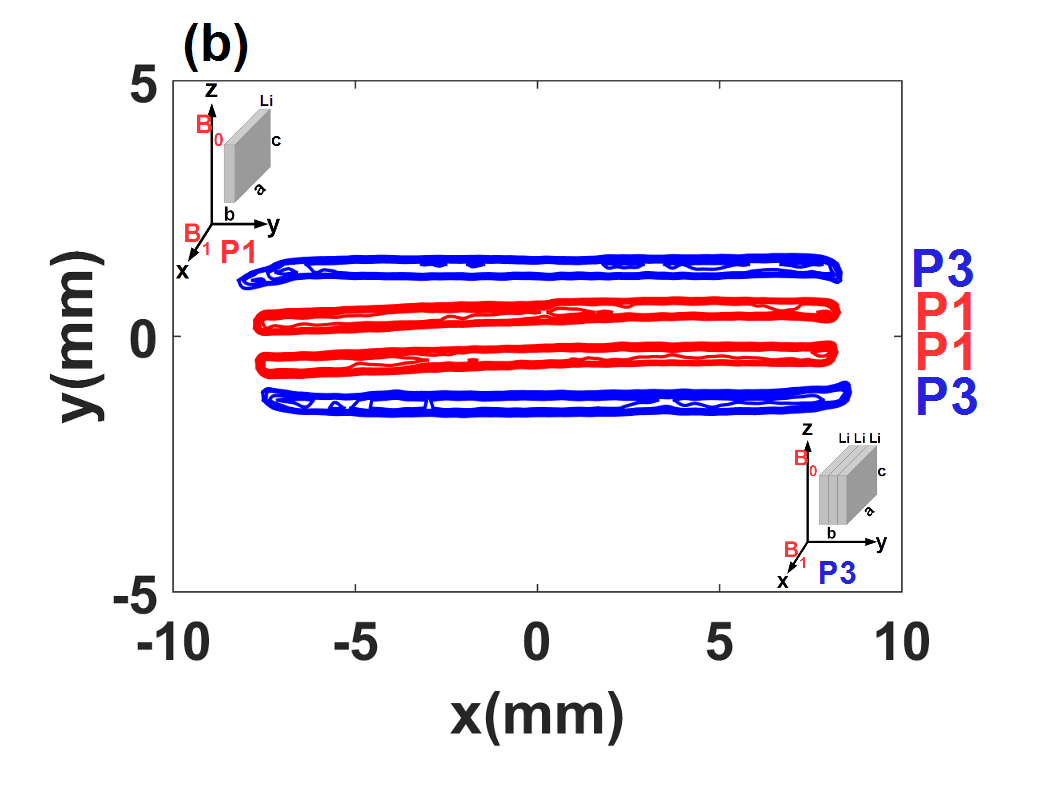}
\label{fig0xy}
\end{figure}

\begin{figure}
\caption{
{\bf \li7 2d CSI($y$) of phantom P3.}\\
Stack plot representation (intensity along the vertical axis).
See also Fig.\ref{fig0csi} and Fig.\ref{fig0phntms}.
Chemical shift along one axis, and image along $y$, comprise the 2 dimensions.
Along $y$, the CS at $\delta_2$ gives rise to a pair of separated bands,
while an extended, low intensity band spanning them is obtained with a CS of
$\delta_1$.
See section \ref{sxn0csi}.
}
\includegraphics[width=9cm]{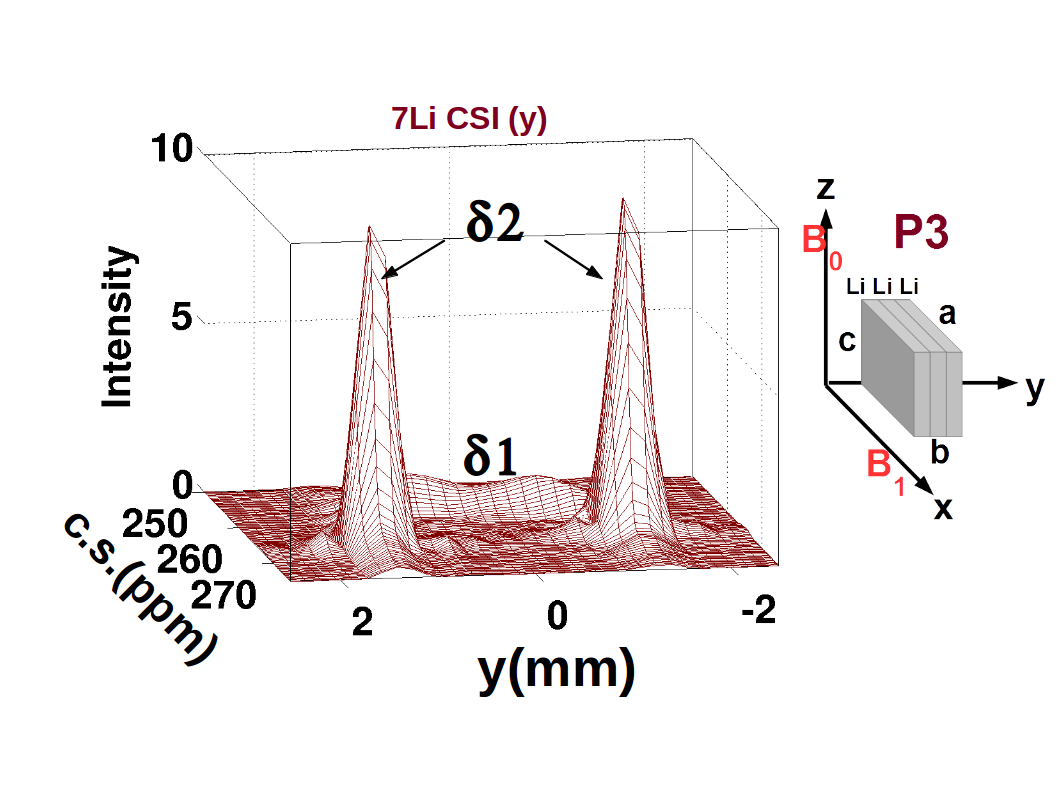}
\label{figSI0csi}
\end{figure}


\begin{figure}
\caption{
{\bf Bulk metal MRI of phantom P3, at mutually orthogonal orientations.}\\
Comparison of \li7 2d MRI stack plots (vertical axis denotes intensity)
of phantom P3 in two different orientations.
\\
{\bf (a)}
2d MRI($yz$) in {\bf \em vertical   } orientation.
{\bf (b)}
2d MRI($zy$) in {\bf \em horizontal } orientation.
The corresponding NMR spectra are shown as insets.

These images declare a virtual {\em dead heat}, between the two orientations,
regarding the relative intensities from the two pairs of $ac$ and $ab$ faces.

This puts to rest the possibility that
the differences in the intensities from the $ab$ and $ac$ faces,
arises due to the differing extents to which the eddy currents (arising from
the gradient switching in MRI experiments) may affect the signals from the
$ab$ and $ac$ faces.

See sections \ref{sxn0formulae},
\ref{subsxn0mthdsMri}
and S\ref{sxnSI0mri2dNtnst}.
}
\includegraphics[width=9cm]{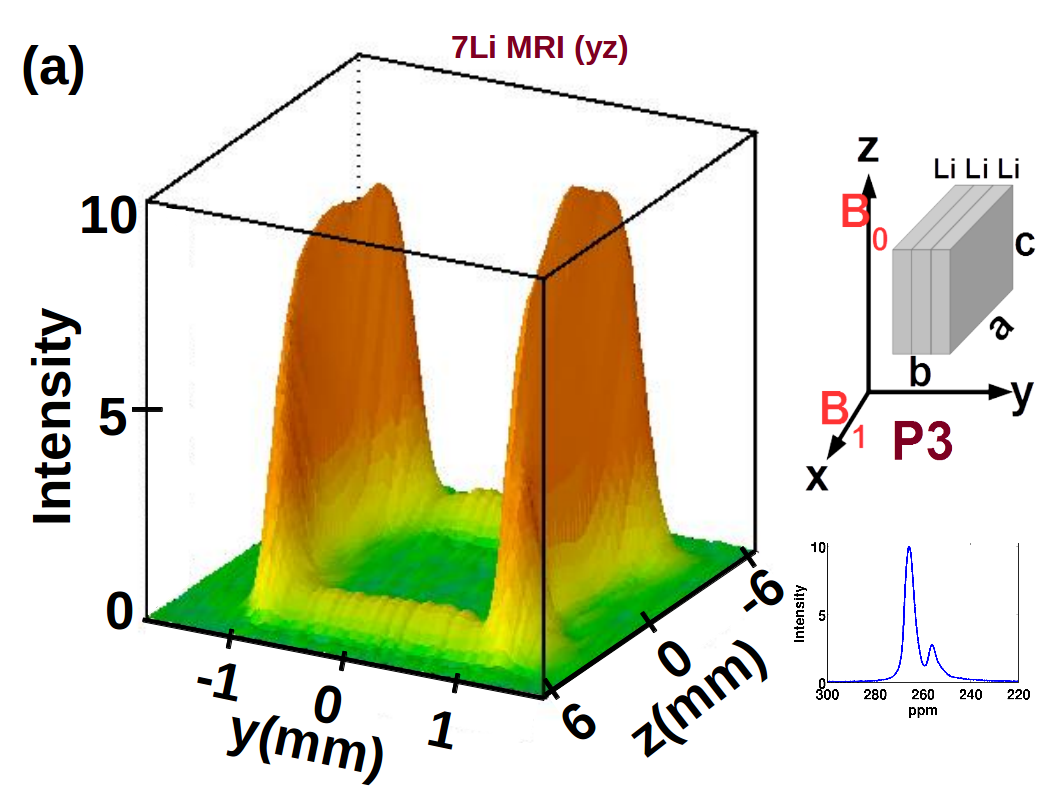}
\includegraphics[width=9cm]{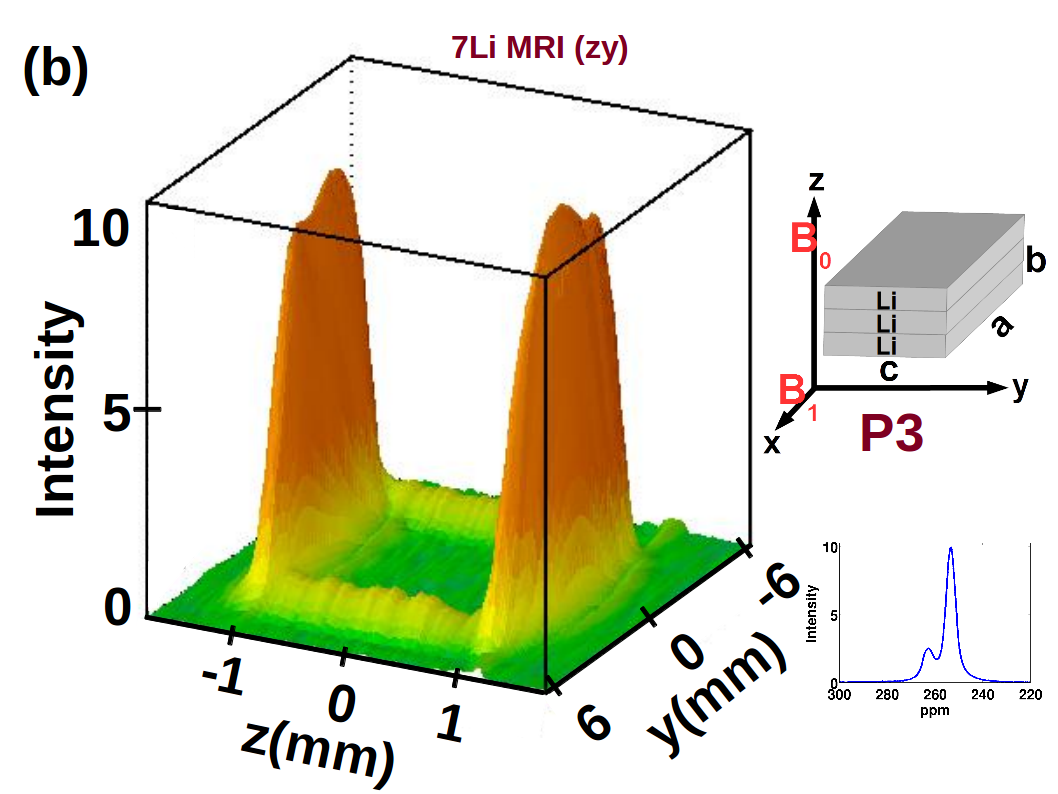}
\label{fig0p3hrzntl}
\end{figure}


\begin{figure}
\caption{
{\bf Resolution, and relative signal intensities from $ab$ and $ac$ faces.}\\
\li7 2d MRI($xy$) stack plots (intensity along the vertical axis) of phantom P3,
at differing resolutions along $y$,
providing experimental verification that the ratio of signals
from $ac$ and $ab$ faces, increases with increasing resolution
(Eq.(\ref{eq:eq0ratioxy}) in section \ref{sxn0formulae}).
\\
{\bf (a), (b), (c)} are respectively at resolutions of 0.5, 0.25 and 0.125
$mm$ along $y$
(Methods section \ref{subsxn0mthdsMri}).
}
\includegraphics[width=16cm]{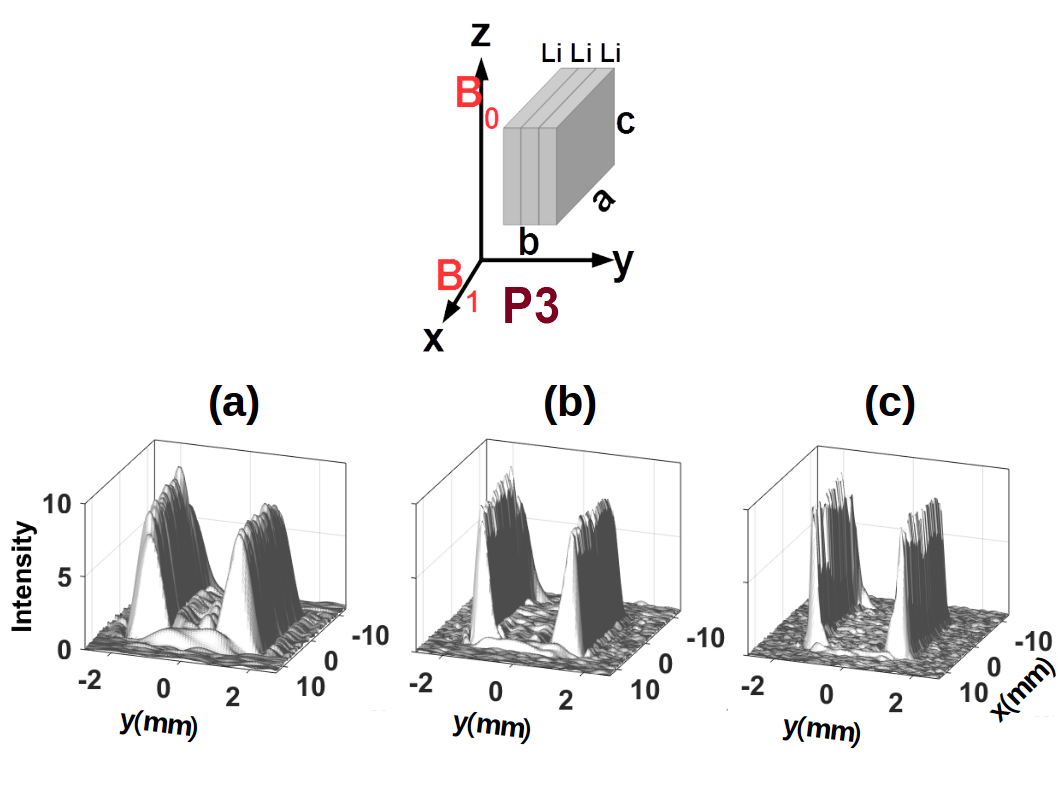} \\
\label{figSI0td20td40td80}
\end{figure}

\begin{figure}
\caption{
{\bf 2d($xy$) slices from MRI($xyz$) of a bulk metal strip.}\\
Visualization of $S_{ab} \neq S_{ac}$
(the signal intensities from $ab$ and $ac$ faces)
{\em via} 2d ($xy$) slices (along $z$) from \li7 3d MRI($xyz$),
of phantom P3 (Fig.\ref{fig0phntms}).\\
{\bf (a)} 3d MRI($xyz$); same data as in Fig.\ref{fig0xyz}. 
{\bf (b)} Slice from the top $ab$ face.
{\bf (c)} Central slice.\\
The slices are displayed as stack plots (intensity along vertical axis).
In a given slice, intensities at all points ($x,y$) are for the same $z$ value
in panel (a).\\
In either slice, the {\em walls} of intensity arise from $ac$ faces.
The {\em plateau} spanning them in panel (b), emanates from the top $ab$
face.\\
See section \ref{sxn0formulae}.
}
\includegraphics[width=16cm]{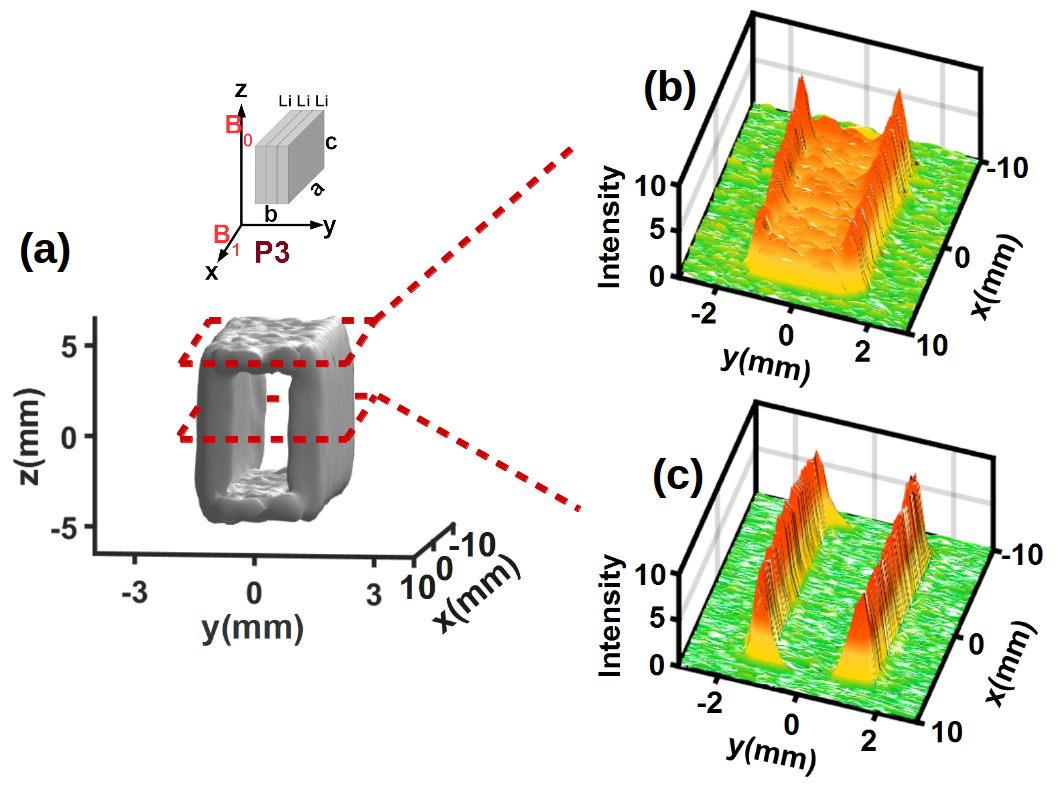} \\
\label{fig0xyzSlc25}
\end{figure}

\end{document}